\documentclass[a4paper,12pt]{article}
\usepackage{graphicx} 
\usepackage{epstopdf} 
\usepackage{amsmath,amsfonts,amssymb,amsthm,mathrsfs,graphicx,float,colordvi,url,wrapfig}
\usepackage{here,color,ulem,enumitem,bm,tikz-cd,blkarray,comment,slashed,cite}
\usepackage{bbm} 
\usepackage[all]{xy} 
\usepackage{tikz} 
\usepackage{scalerel}  
\usepackage[overload]{empheq}
   
\makeatletter

\newcommand{\xdashrightarrow}[2][]{\ext@arrow 0359\rightarrowfill@@{#1}{#2}}
\newcommand{\xdashleftarrow}[2][]{\ext@arrow 3095\leftarrowfill@@{#1}{#2}} 
\newcommand{\xdashleftrightarrow}[2][]{\ext@arrow 3359\leftrightarrowfill@@{#1}{#2}} 
\def\rightarrowfill@@{\arrowfill@@\relax\relbar\rightarrow}
\def\leftarrowfill@@{\arrowfill@@\leftarrow\relbar\relax}
\def\leftrightarrowfill@@{\arrowfill@@\leftarrow\relbar\rightarrow}
\def\arrowfill@@#1#2#3#4{
$\m@th\thickmuskip0mu\medmuskip\thickmuskip\thinmuskip\thickmuskip
\relax#4#1 
\xleaders\hbox{$#4#2$}\hfill
#3$
}
\def\tf{\tilde{f}}

\def\tm{\tilde{m}} 
\def\tb{\tilde{\beta}}
\def\tk{\tilde{\kappa}} 
\def\tD{\tilde{\Delta}}

\def\blue{\textcolor{black}}
\def\Tr{{\rm Tr}}

\def\S4852{S_{\textrm{rad}}^{(\textrm{late})} \big\vert_{\textrm{(\ref{nkre}), (\ref{uygu})}}} 

\newcommand\erase{\bgroup\markoverwith{\textcolor{red}{\rule[.5ex]{2pt}{0.4pt}}}\ULon}

\begin{document}
\renewcommand{\thefootnote}{\fnsymbol{footnote}} 
\begin{titlepage}

\vspace*{0.0mm} 

\begin{center}
{\large \bf 
The Remnant of a Rotating Regular Black Hole 
\\[1.5mm]
from the Generalized Entropy in the Final Stage of Evaporation
}

\vspace*{8.0mm} 
\normalsize
{\large Shingo Takeuchi}
\vspace*{1.5mm}

\textit{
\small 
Faculty of Environmental and Natural Sciences, 
Duy Tan University, Da Nang, Vietnam}\\
\end{center}


\vspace*{-1.0mm} 
\begin{abstract}
\vspace*{+0.0mm} 

We express the generalized entropy~(GE) of the Hawking radiation
in the final stage of an evaporating rotating regular black hole~(BH)
by writing the mass of the BH as $m_{\rm ext}+\alpha$, 
where $m_{\rm ext}$ represents the mass at the extremal limit and $\alpha$ is a parameter.
Generally, entropy is non-negative. 
Based on this fact, we assume that, 
in the GE considered in this study, 
the signs of the contributions 
from the area term and the correction term 
remain unchanged 
throughout the entire evaporation process of the BH. 
Therefore, we regard $\alpha$ 
at which the correction term vanishes  
as its lower bound and determine it. 
As a result, 
we find that 
such a value of $\alpha$ is finite.
Denoting this value by $\alpha_2$,
this result indicates that 
the mass of the BH cannot become smaller than $m_{\rm ext}+\alpha_2$,
which can be interpreted 
as the emergence of a remnant at the final stage of BH evaporation.
The BH considered in this study is a rotating regular BH.
The regularization is motivated by the possibility that 
the fine structure of the central region becomes relevant 
in the final stage of evaporation.
A rotating BH is considered
from the viewpoint of generality.
\end{abstract}

\end{titlepage}

\newpage 

\section{Introduction} 
\label{lcahs}

From a similarity 
between the total derivative of the ADM mass of a black hole~(BH)
and the second law of thermodynamics,
we can infer that thermal radiation is emitted from a BH.
This is known as Hawking radiation~(HR), 
and its temperature is called Hawking temperature~(HT).
\cite{Hawking:1975vcx,Parikh:1999mf,Iso:2006ut} 
can be referred as seminal works for derivation of the HR,
and its existence leads to so-called \textit{the information paradox}, 
which we explain in the following.
\newline

We start with a situation 
in which matter exists. 
It has not yet form a BH. 
We take this situation as the initial state.
Then, we suppose that it forms a BH. 
That BH goes on shrinking due to evaporation by the HR, 
and, normally it is supposed that 
it eventually disappears and only the HR remains. 
We take this situation as the final state.
Then, 
since normally it is assumed that 
the HR always originates from pair-creation 
in the vicinity of the horizon,
the entanglement entropy~(EE) 
between the HR and a BH 
in the final state of evaporation is expected 
to be extremely large 
(where it is assumed that entanglement is maintained even when it crosses the horizon).
However, there is no counter part the HR entangles with at that time.
This is the information paradox.

However, we can expect that 
the information paradox does not arise from the general relationship
$S_{\rm coarse} \ge S_{\rm fine}$,
where $S_{\rm coarse}$ and $S_{\rm fine}$ represent 
the \textit{coarse-grained entropy} and \textit{fine-grained entropy}, 
respectively.
Then, the EE of the HR is classified to $S_{\rm fine}$
and the Bekenstein-Hawking entropy (BH entropy) is classified to $S_{\rm coarse}$. 
Therefore, according to the relation of $S_{\rm coarse}$ and $S_{\rm fine}$, 
the EE of the HR is expected to behave as follows. 
First, it starts from zero as the initial state is a pure state, 
and identified with the EE of the HR.
Therefore, it increases from zero 
in the former stage of evaporation.
But, when it reaches the value of the BH entropy, 
the behavior is switched to the one going along with the BH entropy. 
Therefore, in the later stage of evaporation, 
it decreases along with the BH entropy,
and eventually returns to zero 
when the BH evaporates out and the BH entropy vanishes.
This behavior is referred to as 
\textit{the Page curve}\,\cite{Page:1993wv,Page:2013dx}, 
and the rule to prescribe this switching is 
\textit{the island formula}\,\cite{Penington:2019npb,Almheiri:2019psf} 
(we review this formula in Sec.\,\ref{Sec:ildfml}).
If the Page curve is held, 
the information paradox no longer exists.

Here, the island formula is obtained 
from analysis of the so-called quantum extremal surfaces~(QESs)~\cite{Penington:2019npb,
Almheiri:2019psf,Almheiri:2019hni}. 
The QESs are obtained from the generalized Ryu-Takayanagi (RT) formula, 
which is the RT formula with corrections in the expansion in $G_{\rm N}$~\cite{Hubeny:2007xt,
Lewkowycz:2013nqa,Barrella:2013wja,Faulkner:2013ana, Engelhardt:2014gca}.

Currently, 
we would have a picture 
that the HR always arises from the pair-creations.
If we can change this image
to the one that
the particles inside a BH is solely emitted to the outside 
in the late stage of evaporation,
we can understand the Page curve. 
In fact, in the works cited above~\cite{Hawking:1975vcx,Parikh:1999mf,Iso:2006ut},
the pair-creation is never assumed 
in their analyses deriving HR;
it is a matter of interpretation. 
To confirm this, let us review these studies one by one.

In \cite{Hawking:1975vcx},
the e.v. of 
the particle number operator 
defined on the future null region
is calculated
with the grand state on the past null region
by using the Bogoliubov transformation.
This corresponds to calculating the number of particles 
that originate in the past null region and reach the future null region, 
in terms of particles defined on the future null region.
Its result is interpreted as the HR radiated from the horizon.
In \cite{Parikh:1999mf},
the amplitude of a particle to tunnel through the horizon is analyzed. 
Its square is compared with the Boltzmann factor and the HT is identified.
In \cite{Iso:2006ut}, 
corresponding to a situation 
in which only infalling flux exists 
in the vicinity of the horizon, 
one of the chiral sectors is manually removed 
in the model in the near-horizon region. 
This corresponds to a situation 
in which one of the chiral sectors is absent 
at the classical level.
Then, following two conditions are considered:~1)~the 
general coordinate transformation of the effective 
action obtained from integrating the divergence of 
the energy-momentum tensor should vanish.
2)~There is no flux just on the horizon.
As a result,
the amount of the flux just on the horizon,
which is interpreted as being emitted 
from the horizon via tunneling, 
is determined. 
The result is equivalent 
to the flux of the energy-momentum tensor 
at the temperature given by the HT.

Therefore, 
since pair-creation is not assumed 
in the actual analyses, 
we may modify the picture as described above.
As a result, 
it becomes possible to assume that 
the purifications 
between the newly radiated HR 
and the already radiated HR 
occur
in the late stage of evaporation. 
Consequently, the EE can decrease 
in the late stage of evaporation, 
and the Page curve can be obtained.
\newline

So far, 
the reproductions of the Page curve 
using the island formula
have been carried out
in various BH spacetimes~\cite{Hashimoto:2020cas, 
Wang:2021woy,
Kim:2021gzd, Ahn:2021chg,
Yu:2021rfg, Luongo:2023jyz,
Wang:2024itz, Yu:2025euq
}.
Deriving the island formula is also an important problem, 
which has been addressed 
by analyzing the saddle points associated with the replica Euclidean BH and replica wormhole 
in the gravitational path integral~\cite{Penington:2019kki,Almheiri:2019qdq}.
It has also been performed 
for an evaporating BH~\cite{Goto:2020wnk,Hartman:2020swn} and 
for the black D0-brane system 
described by the BFSS matrix model~\cite{Choudhury:2024hoh}.

On the other hand, 
we focus on the EE 
in the final stage of an evaporating BH. 
The point we emphasize here is the difference 
between our BH and the studies cited above~\cite{Hashimoto:2020cas, 
Wang:2021woy, Kim:2021gzd, Ahn:2021chg, Yu:2021rfg, 
Luongo:2023jyz, Wang:2024itz, Yu:2025euq}. 
These studies consider eternal black holes, 
whereas we consider a shrinking one.
For this purpose, 
we parametrize the BH mass $m_0$ 
as $m_0 = m_{\rm ext} + \alpha$~($m_{\rm ext}$ represents  the mass at the extremal limit),
and express quantities, 
such as the Hawking temperature, the horizon radii,
as expansions in $\alpha$.
Then, using these,
we express the EE of the HR
in the final stage of evaporation.

As a result, 
we find that, 
as $\alpha$ decreases, 
the correction term in the EE of the HR vanishes 
at a certain value of $\alpha$, 
which we denote by $\alpha_2$. 
However, entropy is generally non-negative. 
Therefore, in the EE of the HR, 
the signs of the contributions 
from the area term and the correction term 
are expected 
not to be changed throughout the entire BH evaporation process.
Thus, we regard $\alpha_2$ 
as the lower bound of $\alpha$, 
and determine it.
If $\alpha_2$ is found to be finite, 
this indicates the emergence of a remnant.

Indeed, the HR gives rise to another problem 
in addition to the information paradox mentioned above.
It is currently believed that 
all initial states 
before the formation of a BH
evolve into the same final state 
characterized solely by the HR 
after the BH has completely evaporated. 
This implies that distinct states evolve into the same state, 
which contradicts the unitarity of quantum time evolution.
As possible resolutions to this puzzle, 
corrections to the HR, 
the remnant scenario and 
the firewall proposal~\cite{Almheiri:2012rt,Almheiri:2013hfa}
have been investigated.
Therefore, the remnant 
is one of the important ideas 
for addressing the problem of the unitary evolution in quantum theory.
As a recent concrete study on the emergence of remnants, 
we refer to Ref.~\cite{Easson:2002tg}.
In addition, Refs.~\cite{Aharonov:1987tp,Adler:2001vs,Chen:2014jwq} 
have discussed their fundamental mechanism. 
However, concrete analyses of remnants have not progressed significantly.
This may be because 
the spacetime curvature becomes extremely strong
when the emergence of a remnant is considered. 
In such a situation, 
we investigate the emergence of a remnant 
based on the analysis outlined above.

In the present study, 
we focus on the final stage of an evaporating BH. 
Accordingly, we consider the possibility that 
the structure of the central region of the BH may become relevant.
Therefore, we consider a regular BH (RBH)~\cite{Bardeen,Dymnikova:1992ux}.
In addition, 
since rotating BHs are more generic than non-rotating ones, 
we consider a rotating RBH~\cite{Bambi:2013ufa}. 
\newline

We mention the organization of this study.
In Sec.\,\ref{Sec:ildfml}, 
the island formula is reviewed.
In Sec.\,\ref{Sec:RBH}, 
the rotating RBH spacetime is introduced.
Then, by introducing the parameter $\alpha$ mentioned above,
the horizon radii in the near-extremal regime are obtained.
In Sec.\,\ref{Chap:DRA},
the two-dimensional near-horizon geometry of the rotating RBH spacetime 
considered in this study is obtained. 
Then, from this, the maximally extend spacetime and the surface gravity of our rotating RBH are obtained. 
In Sec.\,\ref{ttjuj}, 
the EE of the HR 
expanded in $\alpha$ 
is obtained.
In Sec.\,\ref{yrtsh}, 
$\alpha_2$ is determined, 
from which the emergence of a remnant 
in the final stage of evaporation is discussed.
In Sec.\,\ref{wh54h}, this study is summarized.
In Appendix\,\ref{App:RBH}, 
the RBH is reviewed.
In Appendix\,\ref{App:Delta4},
the horizon radii are obtained 
by a method different from the one in Sec.\,\ref{Sec:RBH}.
In Appendix\,\ref{App:kukll},
it is shown that
the extremal limit cannot be obtained correctly 
from the condition $r_+=r_-$.
In Appendix\,\ref{App:orhwl},
the assumption used in Sec.\,\ref{mbst}
is derived.
In Appendix\,\ref{ryuyj}, $r_b$ obtained from the extremal condition is presented.

\section{Review of the island formula}
\label{Sec:ildfml} 

We briefly review the island formula, 
which is used in Sec.~\ref{ttjuj} of this study.
Since the Ryu-Takayanagi formula~(RT formula) plays a fundamental role 
in the development of the island formula, 
we begin with the RT formula.
\newline

In considering the RT formula,
we typically consider an AdS space,
and divide its boundary space into the two regions.
Then, we consider a surface 
extending into the bulk space 
from the boundary of these two regions.
We suppose that its area is extremal.
Such an extremal surface is called the quantum extremal surface~(QES), 
and the area of the QES divided by $4G_{\rm N}$ is identified 
with the entanglement entropy~(EE) between the two CFTs 
on the two regions in the boundary space.
(This EE is derived from a path-integral for the gravity in the bulk 
up to order $G_{\rm N}^{-1}$~\cite{Lewkowycz:2013nqa,Gibbons:1976ue} and 
up to order $G_{\rm N}^{0}$~\cite{Faulkner:2013ana,Engelhardt:2014gca}.) 

In the case that the AdS space contains a black hole~(BH), 
it is known that two QESs exist.
If the smaller region shrinks completely to zero~(conversely, if one of the two regions is large), 
one of the QESs becomes a surface enclosing the BH, 
while the other becomes a small surface 
that slightly protrudes from the boundary. 
If it completely shrinks to zero on the boundary space~(conversely, one of the two regions extends to cover the entire boundary space), 
only the QES enclosing the BH remains.
Its area divided by $4G_{\rm N}$ is identified with 
the entropy of the CFT on the boundary space. 
In addition, this is nothing but the Bekenstein-Hawking entropy, 
which represents the entropy associated with the interior of the BH 
and the EE between the interior of the BH and the Hawking radiation~(HR). 
\newline

Now, considering an $AdS_2 \times S^2$ geometry 
obtained from the near-horizon limit of a U(1)-charged extremal BH, 
we consider an eternal BH described by the $AdS_2$ spacetime of this $AdS_2 \times S^2$ geometry.
Then, in order to separate the HR from the interior of the BH, 
we attach a thermal bath~(TB) to each boundary on the left and right sides of the eternal BH, 
where $G_{\rm N}$ is taken to zero so that gravity is absent in the TB.
Actually, thanks to the introduction of the TB, 
the HR is no longer reflected at the boundary, 
and we can identify the entropy of the HR with that of the TB.
We refer to the boundaries on the left and right sides, 
where the TBs are attached, 
as ``the cutoff surfaces'' in the following.

Then, since we are now considering both regions connected by the Einstein-Rosen bridge,
the total Hilbert space for the interior of the BH + the HR 
is expressed by $H_{\rm BH} \otimes H_{\rm TB}$.
Furthermore, denoting the Hilbert space of the CFT on the cutoff surface by $H_{\rm CFT}$, 
since $H_{\rm BH}$ is equivalent to $H_{\rm CFT}$ via the AdS/CFT correspondence mentioned above, 
we can express $H_{\rm total}$ as $H_{\rm CFT} \otimes H_{\rm TB}$.
In order to simplify the description, 
we consider only the right side of the eternal BH in the following.

Therefore, we can express the entropy of the interior of the BH, $S_{\rm BH}$, as
\begin{eqnarray}\label{veri}
S_{\rm BH}
={\rm ext}
\big[
S_{\rm gen}(\Sigma_X)
\big],\quad
S_{\rm gen}(\Sigma_X)
=
{\cal A}(X)/{4G_{\rm N}} + S_{\textrm{field}}(\Sigma_X),
\end{eqnarray}
where 
\begin{itemize}
\item
We gave $S_{\rm BH}$ via the entropy of the boundary CFT, $S_{\rm CFT}$, 
by the holographic manner mentioned above.
The first and second terms represent the leading term and the correction 
at order $G_{\rm N}^0$ 
in the expansion in $G_{\rm N}$, respectively. 
An entropy that includes such a correction is referred to as the ``generalized entropy'', 
which we denoted by $S_{\rm gen}(\Sigma_X)$ as above.

\item
$\Sigma_X$ denotes the region 
extending from the cutoff surface 
into the bulk interior, 
and $X$ denotes its boundary.
The shape of $\Sigma_X$ is determined by extremizing $S_{\rm BH}$ with respect to the position of $X$. 
$\Sigma_X$ and the surface $X$ at this time are 
the so-called entanglement wedge~(EW) and the QES mentioned above, respectively.
Accordingly, ``${\rm ext}[\cdots]$'' means taking the value of $S_{\rm gen}$ 
evaluated on the extremal configuration of $\Sigma_X$.

\item
In $S_{\rm gen}(\Sigma_X)$, 
the first term is determined 
solely by the gravitational action~\cite{Lewkowycz:2013nqa}.
In contrast, the correction term consists of 
contributions from the gravitational and matter fields
in the bulk~\cite{Faulkner:2013ana,Barrella:2013wja,Engelhardt:2014gca}.
However, these can be evaluated separately, 
and the gravitational contribution 
is absorbed into the first term through the renormalization of $G_{\rm N}$. 
As a result, the correction term is ultimately consists solely of the contribution from the fields.  

In conclusion, the first term represents 
the contribution from the gravitational field, 
whereas the second term represents 
the contribution from the bulk fields.

\item
From the fact that $H_{\rm total}$ is given by $H_{\rm CFT} \otimes H_{\rm TB}$,
$S_{\rm BH}$ expresses the EE between the interior of the BH and the HR. 
\end{itemize}

It turns out that two EWs exist:~one extends into the vicinity of the event horizon, 
while the other extends to $r=0$, where its area vanishes and the EW ceases to exist. 
We refer to $S_{\rm BH}$ associated with the former and latter EWs as $S_{\rm non-vanishing}$ and $S_{\rm vanishing}$, respectively.
Therefore, $S_{\rm BH}$ in (\ref{veri}) is finally presented as
\begin{eqnarray}\label{kyri}
S_{\rm BH} = {\rm min}
\{ S_{\rm non-vanishing}, S_{\rm vanishing} \}.
\end{eqnarray}

The Page curve could be obtained if we could evaluate $S_{\rm BH}$ in (\ref{kyri}).
However, the gravitational effect becomes stronger near the BH, 
therefore, the evaluation of $S_{\rm vanishing}$ is practically infeasible.
However, 
if the entropy of the TB could be evaluated, 
since $S_{\rm vanishing}$ represents the EE between the interior of the BH and the TB, 
$S_{\rm vanishing}$ could be obtained from the entropy of the TB. 
We denote the entropy of the TB by $S_{\rm TB}$.s
Therefore, we rewrite (\ref{kyri}) from the perspective of the HR in the following.
\newline

Since the EW associated with $S_{\rm vanishing}$ extends over a constant-time surface 
from one cutoff surface to the cutoff surface on the opposite side in the bulk, 
the complementary space  of this EW is an entire spatial surface at the same time in the TB.
$S_{\rm TB}$ can be given by the entropy of the CFT on this surface.
which is already known to grow linearly with time.

On the other hand, $S_{\rm non-vanishing}$ can be approximately evaluated by identifying it with the area term, 
since the area term dominates and $S_{\textrm{field}}(\Sigma_X)$ is negligible.
This situation holds when viewed from the perspective of $S_{\rm TB}$, 
since the boundaries of the EWs associated with $S_{\rm BH}$ and $S_{\rm TB}$ coincide.

Therefore, we can concretely evaluate $S_{\rm BH}$ from $S_{\rm HR}$, 
which is presented as
\begin{eqnarray}\label{tmrn}
S_{\rm HR} = 
{\rm min}
\big\{ 
{\rm ext}_I
\big[
{\cal A}(\partial I)/{4G_{\rm N}} + S_{\textrm{field}}(I \cup \Sigma_{\rm TB})
\big]
\big\},
\end{eqnarray}
where 
\begin{itemize}
\item
This equation has been obtained 
from the complementary space of $H_{\rm CFT}$ in $H_{\rm CFT} \otimes H_{\rm TB}$ denoted in (\ref{veri}).
\item
$I$ is the so-called ``island'', 
which denotes the region in the bulk that is not included in the EW associated with $S_{\rm CFT}$, 
and $\partial I$ denotes the boundary of the island, 
which is the QES associated with $S_{\rm CFT}$.
$\Sigma_{\rm TB}$ denotes an entire surface 
at a given time in the TB, 
which is the same time as that of the EW of $S_{\rm CFT}$.
$I \cup \Sigma_{\rm TB}$ is the spacial complementary space of the EW associated with $H_{\rm CFT}$.
\item
$S_{\rm non\mbox{-}vanishing}$ and $S_{\rm vanishing}$ in (\ref{kyri}) 
correspond to the cases in which $I$ is finite and vanishes, 
respectively.
\end{itemize} 
In general, 
the island formula~\cite{Almheiri:2019hni,Penington:2019npb,Almheiri:2019psf} 
refers to a formula that determines the EE between the interior of the BH and the HR 
from the perspective of the HR, as in (\ref{tmrn}), and
the behavior of $S_{\rm HR}$ as a function of time gives the Page curve.
For more details, 
see the review paper~\cite{Almheiri:2020cfm,Turiaci:2024cad,Mahajan:2025gfh,Buoninfante:2025gqk,thesis}.
\newline

As such, in the formulation of the island formula,
it is important to consider an AdS spacetime,
since the RT formula is used.
However, in the studies~\cite{Hashimoto:2020cas,
Wang:2021woy, Kim:2021gzd, Ahn:2021chg, Yu:2021rfg,
Luongo:2023jyz, Wang:2024itz, Yu:2025euq},
although the background geometry is not an AdS spacetime,
they assume the existence of an EW and evaluate the EE associated with it.
The present study is also performed assuming the existence of an EW in a spacetime that is not AdS spacetime.

\section{The rotating regular black hole considered in this study}
\label{Sec:RBH} 

In this section, 
we introduce the spacetime 
considered in this study in Sec.\,\ref{Sec:vdsr} 
and comment on the dimensions of the quantities 
used in the paper in Sec.\,\ref{subChap:perpe}.
Then, we obtain the horizon radii 
in the non-extremal and extremal cases 
in Secs.\,\ref{Sec:vius} and \ref{subChap:NHEK2}, 
and show that 
they are continuously connected in Sec.\,\ref{Chap:tdv}.

\subsection{The spacetime considered in this study}
\label{Sec:vdsr} 

In this paper, we consider a rotating regular black hole~(RBH) spacetime given by
\begin{eqnarray}\label{RRBHTD}
ds^2 = g_{tt} dt^2 + 2 g_{t\phi} dt d\phi + g_{\phi\phi}d\phi^2 + g_{rr}dr^2 + g_{\theta\theta}d\theta^2
\end{eqnarray} 
with
\begin{eqnarray}
g_{\mu\nu} &=& 
\begin{pmatrix} 
-\frac{\tD - a^2 \sin^2\theta}{\Sigma} & 0 & 0 & -\big(\frac{r^2+a^2 - \tD}{\Sigma} \big) a\sin^2 \theta \\
0 & {\Sigma}/{\tD} & 0 & 0 \\ 
0 & 0 & \Sigma & 0 \\ 
-\big(\frac{r^2+a^2 - \tD}{\Sigma} \big) a\sin^2 \theta & 0 & 0 & \big( (r^2 + a^2)^2 - \tD a^2 \sin^2 \theta \big) \frac{\sin^2 \theta}{\Sigma}
\end{pmatrix},\nonumber\\*[1.5mm]
\label{SigmatD}
\Sigma \!\! &\equiv& \!\! r^2 + a^2 \cos^2 \theta, \quad
\tD \equiv r^2 -2 \tm r +a^2, \quad
\tm \equiv {m_0 r^3}/{(r^3 + \ell_p^3)}, \quad
a \equiv {J}/{m_0}, \nonumber
\end{eqnarray}
where $\mu,\,\nu = t, r, \theta, \phi$. 
$m_0$ and $J$ denote the ADM mass and the angular momentum of the rotating RBH, respectively.
$\ell_p$ represents a Planck length order quantity, 
which regularizes the singularity in the central region of the BH. 
Indeed, the singularity 
in the Kerr BH is absent
when $\ell_p$ is incorporated into the metric.
We briefly review RBH in Appendix.\,\ref{App:RBH}. 

To confirm the absence of the singularity in our rotating RBH (\ref{RRBHTD}), 
we evaluate $R^2 = R_{\mu \nu \sigma \rho} R^{\mu \nu \sigma \rho}$:
\begin{eqnarray}\label{rtio} 
R_{\mu \nu \sigma \rho}R^{\mu \nu \sigma \rho}
\!\!\! &=& \!\!\!
\frac{192 m_0^2 r^4}{\vartheta^6 \chi^6}
\Big\{
-384 \vartheta ^3 r^6 \chi (2 \ell_p^3+r^3)
+3 \ell_p^6 \chi ^4 (r^3-2 \ell_p^3)^2
+8 \vartheta ^2 r^4 \chi ^2 (53 \ell_p^6+38 \ell_p^3 r^3+9 r^6)
\nonumber\\*[1.5mm]
&& 
\hspace{+17.5mm}
-2 r^2 \chi ^3 
(
52 \ell_p^{12}
+70 \ell_p^9 r^3
+21 \ell_p^6 r^6
+4 \ell_p^3 r^9
+r^{12}
) 
+512 \vartheta^4 r^8
\Big\}
\underset{\theta = \pi/2}{\sim} {1}/{\vartheta^6 \chi^6},
\end{eqnarray} 
where 
$\vartheta \equiv \ell_p^3 + r^3$ and 
$\chi \equiv 2 r^2 + a^2 (1+\cos (2 \theta))$. 
This 
reduces to $R^2$ of the Kerr BH at $\ell_p \to 0$, 
and at this time, a curvature singularity appears 
at $\theta = \pi/2$ and $r = 0$.
In contrast, 
no singularity arises if $\ell_p$ is finite.
\newline

The present study focuses on the final stage of the BH evaporation. 
For this purpose, 
we describe the shrinking BH due to the HR in this regime 
by parameterizing $m_0$ as $m_0 = m_{\rm ext} + \alpha$, 
and decreasing $\alpha$, 
as mentioned in Sec.\,\ref{lcahs}.
For more details regarding these, see Sec.\,\ref{Chap:tdv}.

It will turn out that 
the variables needed to describe the extremal situation
are $m_{\rm ext}$, $r_{\rm ext}$ and $a$, 
whereas the number of equations available to determine them is two.
Therefore, we have two equations for three variables.
In such a situation, we treat $a$ as a parameter.
For more details, see Sec.\,\ref{subChap:NHEK2}.

In the following subsections, 
we obtain the horizon radii 
in the non-extremal and extremal cases, 
and show that they are  continuously connected.
As for the differences between our rotating RBH and the Kerr BH, 
unphysical horizon radii appear, 
and the physical horizon radii differ from those of the Kerr BH 
only by $\ell_p$-dependent corrections. 
These differences do not qualitatively affect the evaporation process. 
Therefore, we assume throughout this paper that 
our rotating RBH continuously shrinks to the extremal limit
without undergoing any phase transition, as does the Kerr BH.

\subsection{The dimensions of the quantities appearing in this study} 
\label{subChap:perpe} 

We employ natural units:~$c=\hbar=1$, 
and $m_0$ is originally given by $G_{\rm N}m_0/c^2$.
As $G_{\rm N}$ is absorbed into $m_0$,
the dimension of $m_0$ is $+1$ in length units. 
In addition, the dimension of $a$ is also $+1$
in length units in natural units.
The dimensions of the remaining quantities are straightforward.

\subsection{The horizon radii in the non-extremal case}
\label{Sec:vius} 

\begin{figure}[H] 
\vspace{-20.0mm} 
\begin{center}
\includegraphics[clip,width=11.7cm,angle=-90]{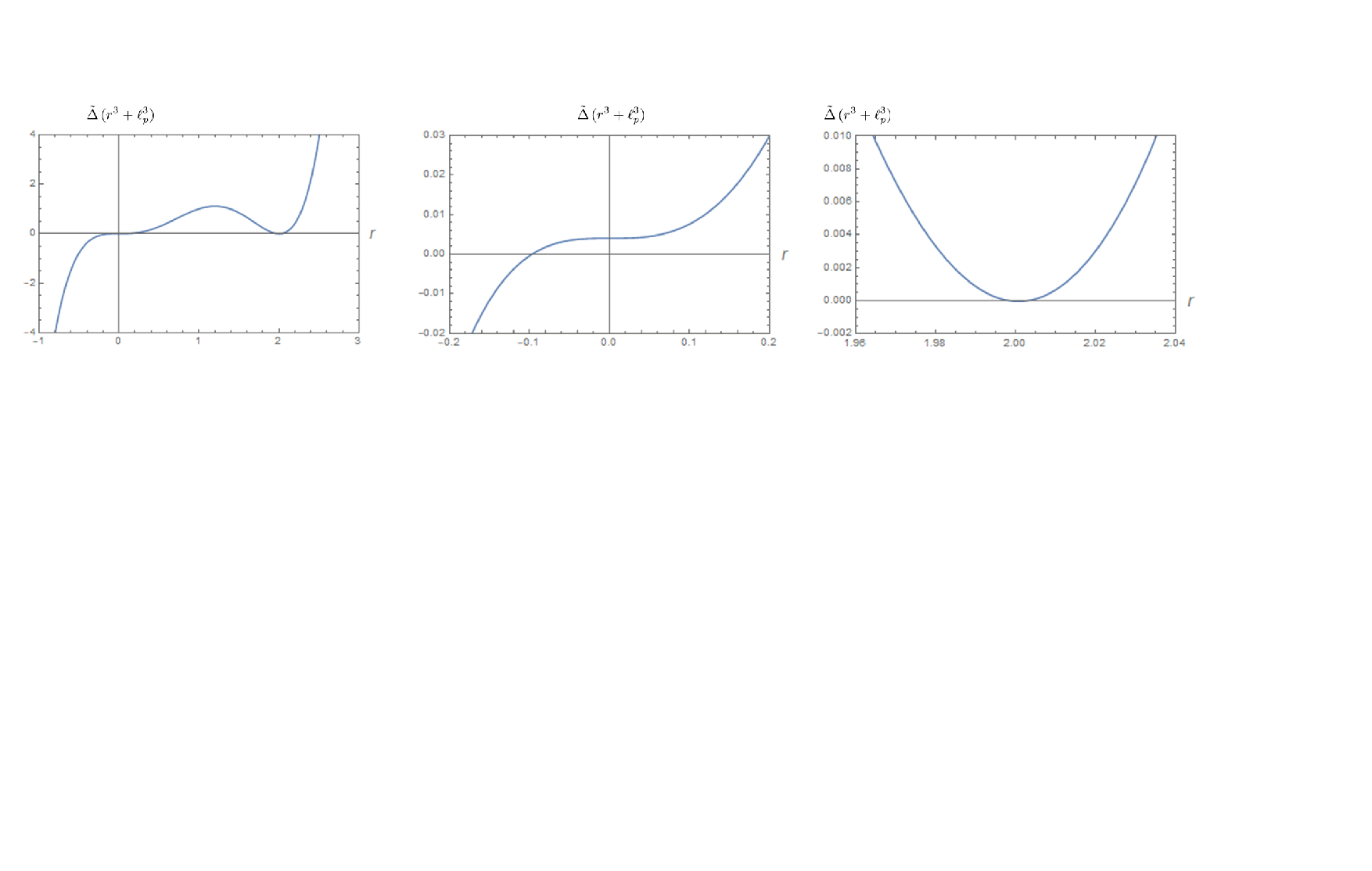} 
\end{center}
\vspace{-74.0mm}
\caption{
Plots of $\tD \, (r^3+\ell_p^3)= 0$ in the non-extremal case with parameters 
$(a, m, \ell_p) = (2,\,4,\,0.1)$.
The right figure shows a magnified view of the left one.
It can be seen that 
$\tD$ admits two positive and one negative real solutions, 
while the remaining two solutions are complex.
} 
\label{wsert1}
\end{figure}

\begin{figure}[H] 
\vspace{-19.mm} 
\begin{center}
\includegraphics[clip,width=11.7cm,angle=-90]{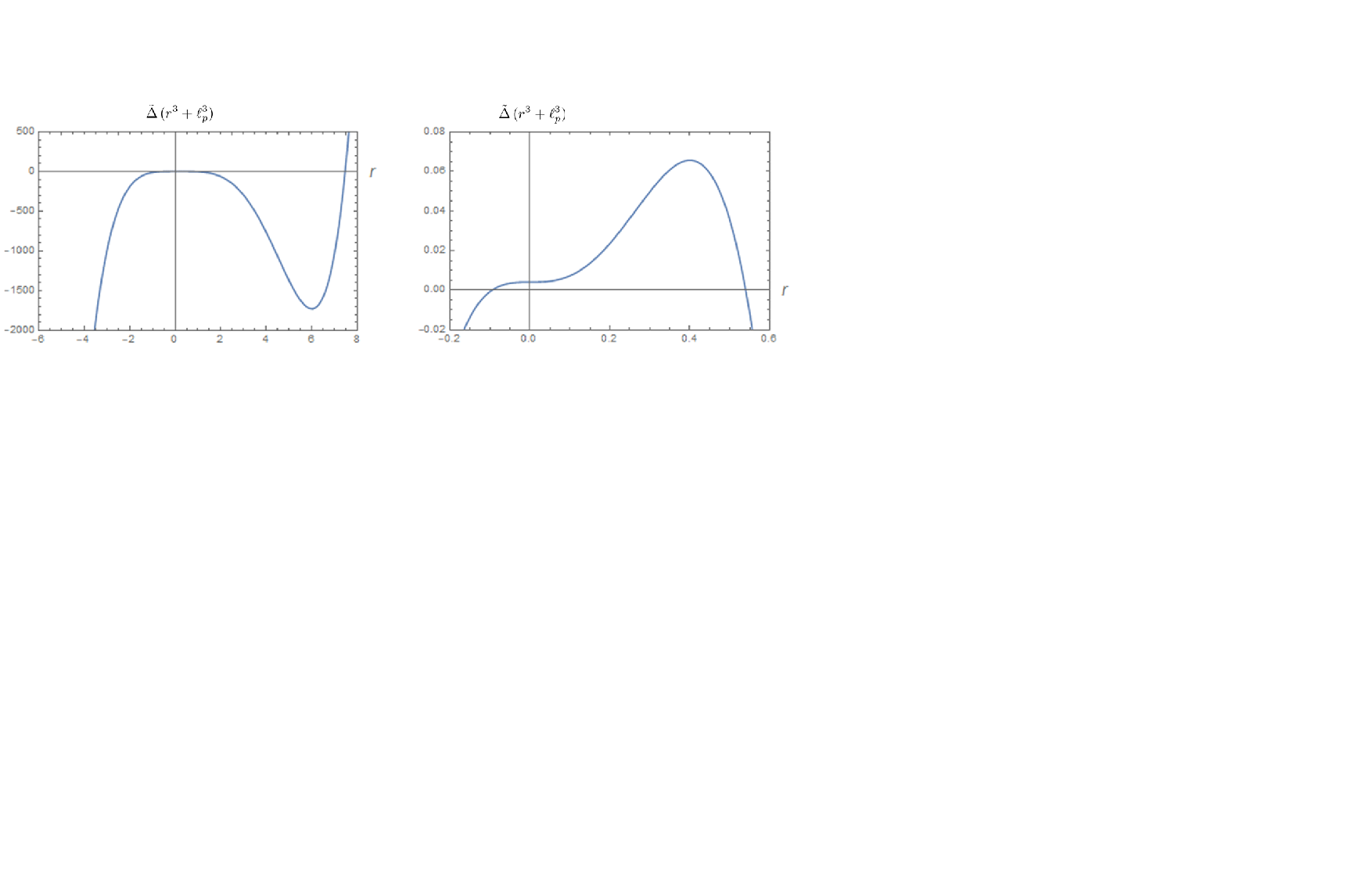} 
\end{center}
\vspace{-74.0mm}
\caption{
Plots of $\tD \, (r^3+\ell_p^3)= 0$ in the extremal case with parameters 
$(a, m_0, \ell_p) = (2,\,2.000282,\,0.1)$.
The middle and right plots show the magnified views of the left one.
It can be seen that 
$\tD$ admits one positive (double root) and one negative solutions. 
} 
\label{wsert2}
\end{figure}

\begin{figure}[H] 
\vspace{-19.0mm} 
\begin{center}
\includegraphics[clip,width=11.7cm,angle=-90]{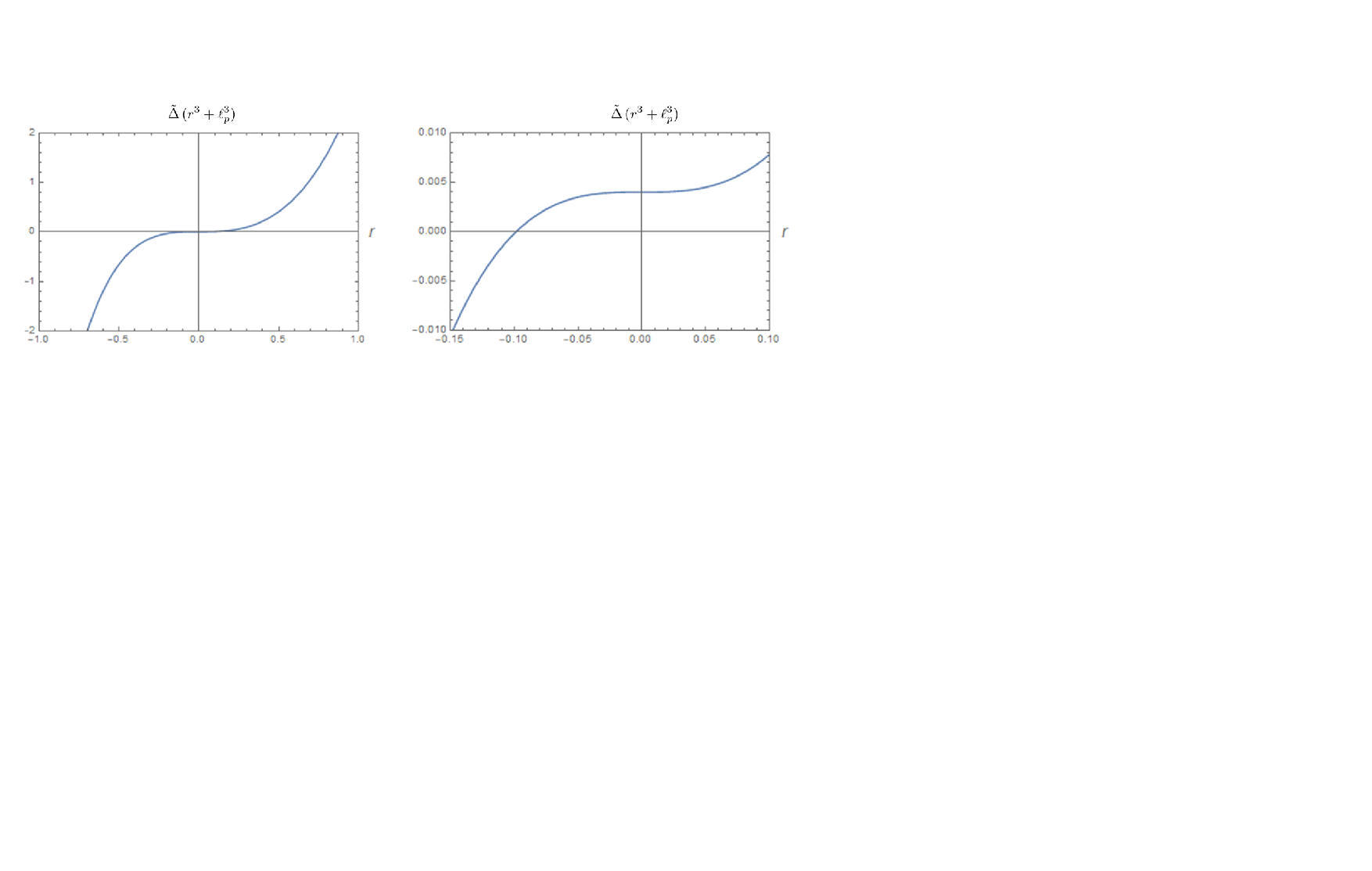} 
\end{center}
\vspace{-74.0mm}
\caption{
Plots of $\tD \, (r^3+\ell_p^3)= 0$ in the case beyond the extremal, 
where $(a,m_0,\ell_p)=(2,\,1,\,0.1)$. 
The right plot is magnified view of the left plot. 
It can be seen that 
$\tD$ admits one negative solution. 
} 
\label{wsert3}
\end{figure}

We obtain the locations of the horizon of (\ref{RRBHTD}). 
For this purpose, we plot $\tD = 0$ 
in the non-extremal, extremal and beyond the extremal cases
in Fig.\,\ref{wsert1},\,\ref{wsert2}~and~\ref{wsert3}, respectively.

$\tD = 0$ is a quintic equation in $r$,
which cannot be solved exactly.
Therefore, 
we assume the form of the solution of $r$ as 
$\gamma_1 + \gamma_2\, \ell_p^3$, 
then substitute this into $\tD = 0$. 
As a result, we can obtain the following expression 
up to $\ell_p^3$ order:
\begin{eqnarray}\label{trnvd}
\tD\vert_{r=\gamma_1+\gamma_2 \,\ell_p^3} =
\alpha^3( \gamma_1^2-2 m_0 \gamma_1+a^2)
+\big\{
a^2+(1+3 a^2 \gamma_2)\,r_0^2 
-8 m_0 \gamma_2 \gamma_1^3
+5 \gamma_2 \gamma_1^4
\big\}\,\ell_p^3 
+\cdots=0.
\end{eqnarray}
Solving this order by order, 
we can obtain the two solutions of $r$ 
up to $\ell_p^3$ order as 
\begin{eqnarray}\label{btyn}
r_\pm 
= 
m_0 \pm \eta
\mp {m_0}/{(m_0\pm\eta)^2\eta}\cdot\ell_p^3
+{\cal O}(\ell_p^4),\quad 
\eta \equiv \sqrt{m_0^2-a^2}.
\end{eqnarray}

Next, assuming the form of the remaining three solutions as 
$\gamma_1 \, \ell_p
+\gamma_2 \, \ell_p^2
+\gamma_3 \, \ell_p^3$.
we substitute this into $\tD=0$ and solve order by order. 
As a result, we can obtain the following three solutions:
\begin{eqnarray}\label{bbile}
&&
r_k
=
e^{i(\pi+2\pi k)/3} \,\ell_p 
+{2m_0}/{3a^2} \cdot e^{2i(\pi+2\pi k)/3} \,\ell_p^2 
-{4m_0^2}/{3a^4} \cdot \ell_p^3 
+{\cal O}(\ell_p^4), \quad 
k=0,1,2, 
\end{eqnarray}
where $e^{i(\pi+2\pi k)/3}\ell_p=-\ell_p$ for $k=1$, 
which corresponds to the negative real solution in Fig.\,\ref{wsert1}.

Thus,
by treating the solutions perturbatively up to $\ell_p^3$ order,
we have solved a quintic equation  $\tD=0$ 
and obtained (\ref{btyn}) and (\ref{bbile}).
These satisfy $\tD=0$
up to $\ell_p^3$ order.
Consequently, $\tD$ in the non-extremal case can be expressed as
\begin{eqnarray}\label{wfyd}
\tD= 
\frac{1}{r^3+\ell_p^3}
(r-r_0)(r-r_1)(r-r_2)(r-r_+)(r-r_-)
\equiv\tD_{\rm non-ext}.
\end{eqnarray}

\subsection{The horizon radii in the extremal case} 
\label{subChap:NHEK2} 

We have plotted $\tD = 0$ in the non-extremal, 
extremal and the beyond-extremal cases 
in Fig.\,\ref{wsert1}--\ref{wsert3}, respectively.
From these figures, we observe the following features 
at the extremal limit:
{\bf 1)}\,two real solutions coalesce into a multiple root,
{\bf 2)}\,a real negative solution is located near $-\ell_p$,
{\bf 3)}\,two of the five solutions are a complex-conjugate pair.
Therefore, we can express $\tD$ in the extremal case as follows:
\begin{eqnarray}\label{tD04}
\tD= 
\frac{1}{r^3 + \ell_p^3}(r-r_{\rm ext})^2(r-r_1)( (r-r_{\rm 2R})^2 + r_{\rm 2I}^2)
\equiv \tD_{\rm ext},
\end{eqnarray}
where 
$r_{\rm ext}$ denotes a multiple real positive root, 
$r_1$ denotes a real negative solution, and
$r_{\rm 2R}$ and $r_{\rm 2I}$ denote 
the real and imaginary parts of the complex-conjugate pair, respectively.
(\ref{tD04}) is a quintic equation 
in which one of the five solutions is a multiple root.
In this subsection, we solve it perturbatively up to $\ell_p^3$ order.
\newline

\begin{figure}[H] 
\vspace{-15.0mm} 
\begin{flushleft}
\includegraphics[clip,width=4.7cm,angle=-90]{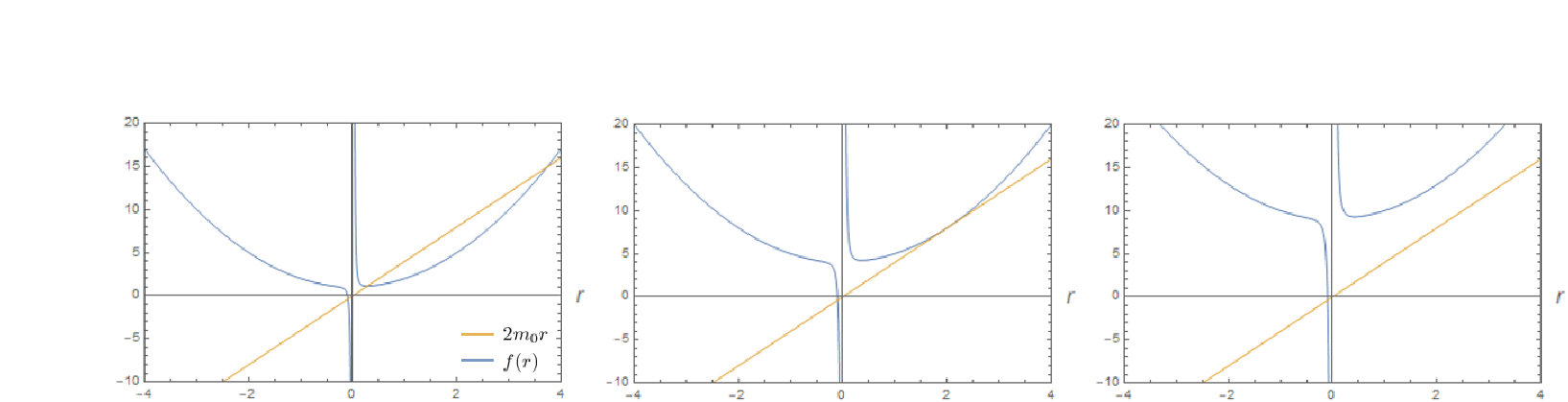} 
\end{flushleft}
\vspace{-3.8mm}
\caption{
Plots of 
$2m_0r$ and 
$f(r)=(r^2+a^2)(1+\ell_p^3/r^3)$, 
which are the functions constituting $\tD = 0$ in (\ref{RRBHTD}),
where $(m_0,a, \ell_p)$ are chosen as $(2,1,0.1)$, $(2,2,0.1)$ and $(2,3,0.1)$ 
in the left, middle and right figures, respectively.
The middle plot corresponds to the extremal case.
} 
\label{ytjr1}
\end{figure}

First, we obtain $r_{\rm ext}$. 
To this end, we rewrite $\tD = 0$ in (\ref{RRBHTD}) as 
$2m_0 r = (r^2+a^2)(1+\ell_p^3/r^3)$, 
and consider the point 
at which $2m_0r$ is tangent to $f(r) \equiv (r^2+a^2)(1+\ell_p^3/r^3)$. 
We plot this behavior in Fig.\,\ref{ytjr1}. 

Since 
a tangent line of a function $f(r)$ at $r=r_{\rm ext}$ can be written as
$f'(r_{\rm ext})(r-r_{\rm ext})+f(r_{\rm ext})$,
we obtain $2m_0r=f'(r_{\rm ext})(r-r_{\rm ext})+f(r_{\rm ext})$. 
From its $r^0$ and $r^1$ orders, 
we obtain 
$0=- f'(r_{\rm ext})\,r_{\rm ext}+f(r_{\rm ext})$ and $2m_0=f'(r_{\rm ext})$, 
which can be explicitly written as
\begin{eqnarray} 
\label{rnewt}
r_{\rm ext}^2 
=a^2+2\,(1/r_{\rm ext}+2a^2/r_{\rm ext}^3)\,\ell_p^3, 
\quad
2\,m_0
=r_{\rm ext}-(1/r_{\rm ext}^2+3a^2/r_{\rm ext}^4)\,\ell_p^3.
\end{eqnarray} 

While (\ref{rnewt}) consists of two equations, 
it involves three variables, $a$, $m_0$ and $r_{\rm ext}$. 
Therefore, one of them must be treated as a parameter. 
In such a situation, 
\begin{eqnarray} \label{vhioe}
\textrm{we take $a$ as a parameter,} 
\end{eqnarray} 
although, in general, 
the angular momentum of a rotating BH decreases as the BH evaporates. 

As such, we solve (\ref{rnewt}) 
for $r_{\rm ext}$ and $m_0$, 
taking $a$ as a parameter.
To this end,
we assume the forms of $r_{\rm ext}$ and $m_{\rm ext}$ up to $\ell_p^3$ order as 
$r_{\rm ext}=a+ \gamma_1 \,\ell_p^3$ and $m_{\rm ext}=a+ \gamma_2 \,\ell_p^3$.
As a result, $r_{\rm ext}$ and $m_{\rm ext}$ are obtained as follows:
\begin{eqnarray}\label{rvsrwh}
r_{\rm ext}=a\, (1+ 3 \, \ell_p^3/a^3 ) +{\cal O}(\ell_p^4), \quad
m_{\rm ext}=a\, (1+ \ell_p^3/a^3 )+{\cal O}(\ell_p^4).
\end{eqnarray} 
From these results, 
$a$ and $r_{\rm ext}$ can be expressed 
in terms of $m_{\rm ext}$ as follows:
\begin{eqnarray}\label{nerh}
a=m_{\rm ext}(1-\ell_p^3/m_{\rm ext}^3)+{\cal O}(\ell_p^4),\quad
r_{\rm ext}=m_{\rm ext}(1+ 2 \, \ell_p^3/m_{\rm ext}^3)+{\cal O}(\ell_p^4).
\end{eqnarray} 
~\newline

Next, we obtain $r_1$.
For this purpose, 
we suppose its form as $r_1= -\ell_p ( 1 + \gamma_1 \,\ell_p/a + \gamma_2 \,\ell_p^2/a^2)$ 
based on the feature {\bf 2)} above. 
As a result, $r_1$ is obtained as follows:
\begin{eqnarray}\label{dsvreh}
r_1= -\ell_p ( 1 -2\,\ell_p/3a+4\,\ell_p^2/3a^2)+{\cal O}(\ell_p^4).
\end{eqnarray}

Lastly, we obtain $r_{\rm 2R}$ and $r_{\rm 2I}$.
The original form of $\tD$ is given in (\ref{RRBHTD}), 
while $\tD$ can be expressed as (\ref{tD04}). 
Therefore, 
we obtain the following relation:
\begin{eqnarray}\label{tretdyh}
-2\,m_{\rm ext} \,r^4+(r^2 + a^2)(r^3+\ell_p^3)
=(r-r_{\rm ext})^2(r-r_1)( (r-r_{\rm 2R})^2 + r_{\rm 2I}^2 ).
\end{eqnarray}
Substituting the solutions of $r_{\rm ext}$ and $r_1$ into this, 
$r_{\rm 2R}$ and $r_{\rm 2I}$ can be obtained as follows:
\begin{eqnarray}\label{terbe}
r_{\rm 2R} 
= \ell_p 
( 
{1}/{2} 
- \ell_p/3a 
- 4 \ell_p^2/ 3 a^2 
)
+{\cal O}(\ell_p^4), \quad 
r_{\rm 2I} 
= \ell_p ( 
\sqrt{3}/2 
+ \ell_p / \sqrt{3}a
)
+{\cal O}(\ell_p^4).
\end{eqnarray}

We can verify that 
the solutions obtained in this subsection satisfy $\tD=0$ 
up to $\ell_p^3$ order.
One may attempt to obtain these solutions 
based on the condition $r_+=r_-$. 
However, this approach does not work.
We discuss this issue in Appendix\,\ref{App:kukll}. 

\subsection{Continuity between the non-extremal and extremal cases}
\label{Chap:tdv}

In this subsection, 
we show that 
the horizon radii 
in the non-extremal and extremal cases 
obtained in Secs.\,\ref{Sec:vius} and \ref{subChap:NHEK2}
continuously connect each other.
For this purpose, 
we express $m_0$ as 
\begin{eqnarray}\label{owros}
m_0 = m_{\rm ext} + \alpha,
\end{eqnarray} 
and parametrize $m_0$ by $\alpha$, 
where $\alpha$ is a parameter and $m_{\rm ext}$ has been obtained in (\ref{rvsrwh}).
The limit $\alpha \to 0$
corresponds to the extremal limit.
Then, we express the non-extremal horizon radii $r_\pm$, $r_{0,1,2}$ 
in the near-extremal regime as
\begin{subequations}\label{wred}
\begin{align}
\label{wred1}
\cdot \quad \!\!
r_\pm \vert_{\rm (\ref{owros})}
=& \,\,
a \pm \sqrt{2 \alpha a}
+{\cal O}(\alpha)
+a^{-2}
\big(
3
\mp {2\sqrt{2 \alpha}}/{a^{1/2}}
+{\cal O}(\alpha)
\big) \,\ell_p^3
+{\cal O}(\ell_p^4)
\nonumber\\*[1.5mm]
=& \,\,
r_{\rm ext}+
( 
\pm \sqrt{2a} \mp {2\sqrt{2}}\,\ell_p^3/{a^{5/2}}
)\sqrt{\alpha}+\cdots,
\\*[1.5mm]
\label{wred2}
\cdot \quad \!\!\!
\hspace{0.9mm}
r_1\vert_{\rm (\ref{owros})}
=& 
-\ell_p
+\big({2}/{3 a}+{2 \alpha}/{3 a^2}+{\cal O}(\alpha^2)\big)\,\ell_p^2
-\big({4}/{3 a^2}+{8 \alpha}/{3 a^3}+{\cal O}(\alpha^2)\big)\,\ell_p^3
+{\cal O}(\ell_p^4)
\nonumber\\*[1.5mm]
=& \,\, 
r_1^{\textrm{(ext)}}
+2\alpha/3 a^2\cdot 
({\ell_p^2}-{4\ell_p^3}/{a})
+\cdots,
\\*[1.5mm]
\label{wred3}
\cdot \quad \!\!\!
\hspace{0.9mm}
r_{0}\vert_{\rm (\ref{owros})}
=& \,\,
(-1)^{1/3}\ell_p
+\frac{2}{3a}\cdot (-1)^{2/3}
\big(
1+\frac{\alpha}{a}+{\cal O}(\alpha^2)
\big)\,\ell_p^2
-\frac{4}{3a^2} \cdot
\big(
1+\frac{2\alpha}{a}+{\cal O}(\alpha^2)
\big)\,\ell_p^3+{\cal O}(\ell_p^4)
\nonumber\\*[1.5mm]
=& \,\, 
r_{\rm 2R}^{\textrm{(ext)}} + i r_{\rm 2I}^{\textrm{(ext)}}
+{\alpha}/{a^2}\cdot
\big(
(-1/3+{i}/{\sqrt{3}})\,\ell_p^2
-{8\ell_p^3}/{3a}
\big)+\cdots,
\\*[1.5mm]
\label{wred4}
\cdot \quad \!\!\!
\hspace{0.9mm}
r_2\vert_{\rm (\ref{owros})}
=& \,\,  
(r_0\vert_{\rm (\ref{owros})})^\ast,
\end{align} 
\end{subequations} 
where 
\begin{itemize}

\item
$r_\pm$, 
$r_{1,0}$ and $r_{\rm ext}$
are given in (\ref{btyn}), (\ref{bbile}) and (\ref{rvsrwh}), respectively.
Meanwhile,
$r_1^{\textrm{(ext)}}$, $r_{\rm 2R}^{\textrm{(ext)}}$ and $r_{\rm 2I}^{\textrm{(ext)}}$
denote
$r_1$ in (\ref{dsvreh}),
$r_{\rm 2R}$ and $r_{\rm 2I}$ in (\ref{terbe}),
respectively.

\item
$a$ is a parameter as mentioned in (\ref{vhioe}), 
and 
we assume that 
these expansion forms remain valid 
up to $\alpha=0$~(this is equivalent to assuming that the adiabatic approximation holds up to the extremal limit).

\item
We see that the horizon radii in the non-extremal case coincide with
those in the extremal case in the extremal limit.

\item
It may appear that 
a branch arises from $\sqrt{\alpha}$. 
However, since $\alpha$ is supposed to be real in this study, 
no such branch arises 
in this study.

\item
$r_\pm \vert_{m_0=m_{\rm ext}}$ (the direct extremal limit of $r_\pm$ without the parametrization of $\alpha$) 
do not coincide with $r_{\rm ext}$.
This is because 
$\eta$ in (\ref{btyn})
ceases to hold the power-series expansion with respect to $\ell_p^3$ 
when $m_{\rm ext}$ is directly substituted into $m_0$\footnote{
Actually, 
$\eta \vert_{m_0=m_{\rm ext}}=\sqrt{2/a}\cdot \ell_p^{3/2}+\cdots$, 
while, 
$\eta\vert_{m_0=m_{\rm ext}+\alpha}
=\sqrt{2a\alpha}+(1/2\sqrt{2\alpha}a^{3/2}+3\sqrt{\alpha}/8\sqrt{2}a^{5/2}+\cdots)\ell_p^3+\cdots$.
}. 
\item
$(-1)^{1/3}
=1/2+\sqrt{3}i/2
$ has been used in (\ref{wred3}).
\end{itemize}

We can see that 
$\tD_{\rm non-ext}$ in 
(\ref{wfyd}) 
can be expressed in the near-extremal regime as
\begin{eqnarray}\label{nklsd}
\tD_{\rm non-ext} 
\vert_{\rm (\ref{owros})}
\!\!\! &=& \!\!\!
(a-r)^2
-2 \alpha r 
+{\cal O}(\alpha^{3/2})
+{\cal O}(\alpha^{3/2})\,\ell_p^2
+(
{2 a}/{r^2}
-{2 r}/{a^2}
+{2 \alpha}/{r^2}
+{\cal O}(\alpha^{3/2})
)\,\ell_p^3
+ {\cal O}(\ell_p^4)
\nonumber\\*[1.5mm]
\!\!\! &=& \!\!\!
\tD_{\rm ext}\vert_{\textrm{(\ref{rvsrwh}),\,(\ref{dsvreh}),\,(\ref{terbe})}}
+2 \alpha (-r+{\ell_p^3}/{r^2})
+\cdots,
\end{eqnarray} 
where $\tD_{\rm ext}$ is given in (\ref{tD04}).
From the results (\ref{wred}) and (\ref{nklsd}),
we see that 
the non-extremal case is continuously connected to the extremal case 
in our rotating RBH.

We can verify 
that (\ref{nklsd}) coincides with 
the near-extremal regime of 
$\tD$ given by the original expression in (\ref{RRBHTD}) 
to the order considered in the analysis
as follows:
\begin{eqnarray}\label{rrykr}
\textrm{$\tD$ in (\ref{RRBHTD})} 
\vert_{\textrm{(\ref{owros})}}
\!\! &=& \!\!
(a-r)^2
-2 \alpha r 
+{\cal O}(\alpha^2)
+(
{2 a}/{r^2}
-{2 r}/{a^2}
+{2 \alpha}/{r^2}
+{\cal O}(\alpha^2)
)\,\ell_p^3 
+ {\cal O}(\ell_p^4)=\textrm{(\ref{nklsd})}.
\end{eqnarray}
Although $\tD_{\rm ext}$ and the original $\tD$ represent the same quantity, 
their expressions are  different. 
Therefore, this coincidence is not always obvious.

\section{The maximally extended spacetime}
\label{Chap:DRA}

The quantum extremal surfaces~(QES) 
in the island formula 
are typically discussed 
in the maximally extended spacetime 
for the non-rotating BHs~\cite{Hashimoto:2020cas, 
Wang:2021woy,Kim:2021gzd, Ahn:2021chg,
Yu:2021rfg, Luongo:2023jyz}
and rotating BHs~\cite{Wang:2024itz,Yu:2025euq}.
Therefore, in this section,
we construct the maximally extended spacetime~(MES)
associated with our rotating RBH.

However, the MESs of non-rotating BHs are usually geodesic hypersurfaces, 
whereas those of rotating BHs are not 
for arbitrary $\theta$ 
except at $\theta=0, \pi$.
This is because
the $(t,r)$ and angular sectors are not decoupled
for $\theta \not= 0,\pi$ in rotating BHs;
as a result, 
to construct the MES of a rotating BH,
we must rewrite the metric into the co-rotating frame,
and then fix $\theta$ and $\tilde{\phi}$~($\tilde{\phi}$ denotes the $\phi$ direction in the co-rotating frame).
However, since the $(t,r)$ section 
specified by $\tilde{\phi}={\rm const.}$ 
is not a geodesic hypersurface except at $\theta = 0,\pi$, 
it is not appropriate to discuss configurations of QESs in the MES of rotating BHs.

In such a situation, Refs.~\cite{Wang:2024itz,Yu:2025euq}
extract the near-horizon geometries from field theories near the horizon,
then employ the MESs obtained from these near-horizon geometries 
as those where they consider the configurations of MESs.
 
In this study, 
we follow the same approach. 
Accordingly,
we analyze the effective action of a scalar field 
in the near-horizon region, 
from which
we extract the two-dimensional near-horizon geometry 
in Sec.\,\ref{Chap:espog}.
From this, 
we derive the surface gravity 
in Sec.\,\ref{Chap:rgtr}, 
and then construct the maximally extended spacetime 
in Sec.\,\ref{vber}.

\subsection{The effective action in the near-horizon region}
\label{Chap:espog}

We consider the following scalar theory: 
\begin{eqnarray}\label{scalar}
S= 
\frac{1}{2} \, 
\int_{-\infty}^\infty \! dt
\int \! dr
\int \! d\Omega \, \sqrt{-g} \, 
( 
g^{\mu\nu} \partial_\mu \phi^{\ast} \partial_\nu \phi 
+ {\cal L}_{\rm mass} + {\cal L}_{\rm int} 
)
\equiv
\int \! d^4x \, \sqrt{-g} \,{\cal L}, 
\end{eqnarray}
where 
$\int dr=\int_{r_+}^\infty dr$ 
for the non-extremal case, and
$\int_{r_{\rm ext}}^\infty dr$
for the extremal case.
$g^{\mu\nu}$ is the inverse metric of (\ref{RRBHTD}).
To maintain generality, 
we have included the mass and interaction terms in the above form.
We separate the near-horizon contribution of (\ref{scalar}) as 
\begin{eqnarray}\label{seart}
S = 
\underbrace{
\int_{r_+}^{r_+ + \delta} \! dr \int \! d^3x \, \sqrt{-g} \, {\cal L}}_{\equiv S_{\rm nh}}
+\int_{r_+ + \delta}^\infty \! dr \int \! d^3x \, \sqrt{-g} \, {\cal L},
\end{eqnarray}
where $\delta$ is an infinitesimal parameter
specifying the near-horizon region. 
Based on the general fact that
${\cal L}_{\rm mass}$ and ${\cal L}_{\rm int}$ become negligible 
compared with the kinetic term 
in the near-horizon region 
due to the strong gravitational effect.
Therefore, we can consider $S_{\rm nh}$ solely by the kinetic term as follows:
\begin{eqnarray}\label{ryty}
S_{\rm nh}
=
\frac{1}{2}\int d^4x \, 
\sqrt{-g} \, g^{\mu\nu} \partial_\mu \phi^\ast \partial_\nu \phi 
\!\!\! &=& \!\!\! 
\frac{1}{2}
\int d^4x \sqrt{-g}\,
\Big[ 
-\phi^\ast \Big\{ 
(
\sqrt{-g})^{-1}\partial_\theta(\sqrt{-g}g^{\theta\theta} \partial_\theta)
+g^{\phi\phi} \partial^2_\phi 
\Big\} \phi 
\\*[1.5mm]
&& \hspace{25.5mm}
- \, \phi^\ast \Big\{ 
g^{tt} \partial_t^2 
+ (\sqrt{-g})^{-1}\partial_r(\sqrt{-g}g^{rr}\partial_r)
+ 2g^{\phi t} \partial_\phi \partial_t 
\Big\} \phi \, 
\Big].
\nonumber 
\end{eqnarray} 
As a result, $S_{\rm nh}$ can be treated as an isolated system.

Denoting $S_{\rm nh}$ in (\ref{ryty})
as $\frac{1}{2} \int d^4x({\cal B}_1+{\cal B}_2)$, 
${\cal B}_1$ and ${\cal B}_2$ are expressed as 
\begin{subequations}\label{tuyld}
\begin{align} 
\label{GIIIAngleP1} 
{\cal B}_1
=& 
-\phi^\ast \,
\Big\{
\partial_\theta (\sin \theta\,\partial_\theta)
-{a^2/{\tD}\cdot\sin \theta}\,\partial_\phi^2 \,
\Big\}
\phi
\rightarrow
\phi^\ast \,
{a^2/{\tD} \cdot \sin \theta}\,\partial_\phi^2 \,
\phi+{\cal O}((r-r_+)^0),
\\*[1.5mm]
\label{GIIIAngleP2}
{\cal B}_2
=&
-\sqrt{-g}\,\phi^\ast g^{tt} 
\Big\{ 
( \partial_t + {g^{\phi t}}/{g^{tt}}\cdot\partial_\phi )^2 
-({g^{\phi t}}/{g^{tt}}\cdot\partial_\phi)^2
+ (g^{tt} \sqrt{-g})^{-1}\,\partial_r(\sqrt{-g}\,g^{rr}\partial_r) 
\Big\} \phi 
\nonumber\\*[1.5mm]
\rightarrow& \,
\phi^\ast \, \sin \theta \,
\Big\{ 
{(a^2+r^2)^2}/{\tD} \cdot ( \partial_t + {a}/{a^2+r^2}\cdot\partial_\phi )^2 
- {a^2}/{\tD} \cdot\partial_\phi^2 
- \partial_r (\tD\,\partial_r)
\Big\} \phi+{\cal O}((r-r_+)^0),
\end{align} 
\end{subequations}
where 
we denoted $\tD_{\rm non-ext}$ as $\tD$ for simplicity.
Accordingly,
\begin{eqnarray}\label{jjsrt}
S_{\rm nh} \to \frac{1}{2} \int d^4x \,
\phi^*
\sin \theta \,
\Big\{ 
{(a^2+r^2)^2}/{\tD}\cdot(\partial_t + {a}/{(a^2+r^2)}\cdot\partial_\phi)^2 
- \partial_r ( \tD \,\partial_r ) 
\Big\} \phi + {\cal O}((r-r_+)^0),
\end{eqnarray}
where ``$\to$'' means retaining the leading terms in $\tD \sim 0$ in $r \sim r_+$. 
If we consider the extremal case, 
it is given 
by $\tD_{\rm ext}$ and $r_+$ are replaced with $r_{\rm ext}$.

Now, we expand the scalar field 
using the spherical harmonics $Y_{lm}$ as
\begin{eqnarray}\label{fstr}
\phi(t,r,\theta,\phi) 
\!\! &=& \!\! 
\sum_{l=0}^\infty \sum_{m=-l}^l \varphi_{lm}(t,r) \, Y_{lm}(\theta, \phi).
\end{eqnarray}
Then, by integrating out the angle directions, 
we can give $S_{\rm nh}$ in (\ref{jjsrt}) as follows:
\begin{eqnarray}\label{MetricsType}
S_{\rm nh} 
\!\!\! &=& \!\!\! 
- \frac{1}{2}\,\sum_{lm}
\int dt dr \, \Phi \, \varphi_{lm}^* \,\big( 
g_{\text{nh}}^{tt} ( \partial_t - im A_t )^2 
+ \partial_r (g_{\text{nh}}^{rr} \, \partial_r
)\big) \,\varphi_{lm}, 
\\*[1.5mm]
g_{{\rm nh},\, ij}
\!\!\! &\equiv& \!\!\! 
\left(
\begin{array}{cc}
-\tf,\ & 0 \\[1.0mm]
0 & \tf^{-1} \\
\end{array}
\right),\quad 
\tf 
\equiv \frac{\tD}{a^2+r^2},
\quad
\tD=\left\{
\begin{array}{ll}
\tD_{\rm non-ext} & \textrm{for the non-extremal case given in (\ref{wfyd}),} \\[1.5mm] 
\tD_{\rm ext} & \textrm{for the extremal case given in (\ref{tD04}),} \\
\end{array}
\right.
\nonumber
\\*[1.5mm]
(A_t,\, A_r) \!\!\! &\equiv& \!\!\! (-{a}/({a^2+r^2}),\, 0 ),\quad 
\Phi \equiv a^2+r^2, 
\nonumber
\end{eqnarray}
where 
$
-i \, \partial_\phi Y_{lm}
= m \, Y_{lm}$ and
$\int d\theta d\phi \, \sin \theta \, Y^\ast_{l'm'}Y_{lm}=\delta_{l'l}\,\delta_{m'm}$. 
We also used the relation
$
\frac{1}{r^2+a^2}\partial_r (\tD \partial_r) \to 
\partial_r ( \frac{\tD}{r^2+a^2} \partial_r)
$.
From the result (\ref{MetricsType}), 
we see that $S_{\rm nh}$ is effectively given
by a collection of scalar field theories 
labeled by $(l,m)$ 
on the two-dimensional spacetime $(t,r)$ 
with the metric $g_{{\rm nh},\, ij}$.

\subsection{The surface gravity}
\label{Chap:rgtr}

From (\ref{MetricsType}), 
the surface gravity of our rotating RBH (\ref{RRBHTD}) 
is obtained as
\begin{eqnarray}\label{thyrthrt}
\tk
=
\left\{
\begin{array}{ll}
\displaystyle{ 
\frac{r_+ - r_-}{2 (a^2+r_+^2)}
+\frac{ m_0 (r_+-r_-) (a^2+2 m_0 r_+)}{a^4 r_+^2 (a^2+r_+^2)}\ell_p^3
+ {\cal O}(\ell_p^4)
} & \hspace{-2mm} \equiv \tk_{\textrm{non-ext}}, 
\\[3.0mm] 
0 & \hspace{-2mm} \equiv \tk_{\textrm{ext}}, 
\end{array}
\right.
\end{eqnarray}
where $\tk_{\textrm{non-ext}}$ is the surface gravity at $r=r_+$, 
and we have used the relation:~$\tk =2 \pi / \tb =\partial_r \tf /2 \vert_{r=r_+}$ 
($\tb$ denotes the inverse Hawking temperature of our rotating RBH). 
By substituting the expressions for $r_\pm$ from (\ref{wred}), 
$\tk_{\rm non-ext}$ 
in the near-horizon region is given by
\begin{eqnarray}\label{eusrtr}
\tk_{\rm non-ext} \vert_{r=r_\pm \,\textrm{in (\ref{wred})}}= 
({\sqrt{\alpha}}/{\sqrt{2}a^{3/2}}+{\cal O}(\alpha))
+
({\sqrt{\alpha}}/{\sqrt{2}a^{9/2}}+{\cal O}(\alpha))\,
\ell_p^3+ {\cal O}(\ell_p^4).
\end{eqnarray}
We see that
$
\tk_{\rm non-ext} 
\vert_{
r=r_\pm~\! \textrm{in (\ref{wred})}
}
\to \tk_{\rm ext}
$
at $\alpha \to 0$.

For $a/m_0 \ll 1$, 
$\tk_{\rm non-ext}$ 
in (\ref{thyrthrt}) is expressed as
\begin{eqnarray}\label{euftr}
\tk_{\rm non-ext}\vert_{a/m_0 \ll 1}= 
\frac{1}{4m_0}
\Big(1-\frac{a^2}{4m_0^2}+{\cal O}((\frac{a}{m_0})^4)\Big)
-\frac{1}{16m_0^2}
\Big(1+\frac{3a^2}{2m_0^2}+{\cal O}((\frac{a}{m_0})^4)\Big)\ell_p^3
+{\cal O}(\ell_p^4).
\end{eqnarray}
From this result, it follows that
$\tb^{-1}$ diverges as $m_0 \to 0$.

In this study, 
$a \sim m_0$ 
in the near-extremal regime, 
as can be seen in (\ref{nerh}). 
Therefore, 
$\tk_{\rm non-ext}$ behaves according to (\ref{eusrtr}) 
in the near-extremal regime and vanishes 
in the extremal limit. 

\subsection{The maximally extended spacetime} 
\label{vber}

\begin{figure}[H] 
\vspace{-22mm} 
\hspace{-0mm} 
\begin{center}
\includegraphics[clip,width=10.5cm,angle=-90]{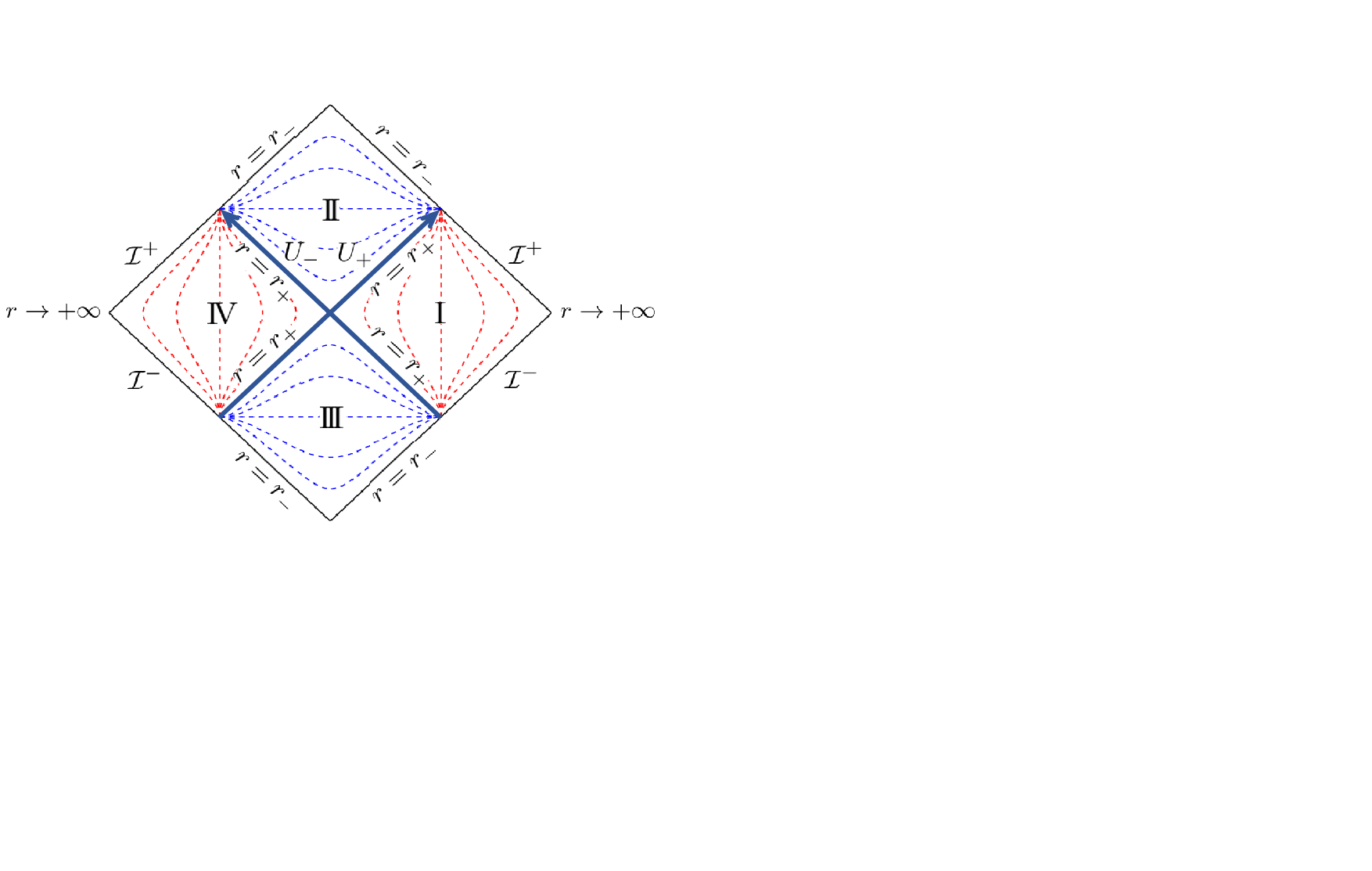} 
\end{center}
\vspace{-45.0mm}
\caption{
The Penrose diagram
of the maximally extended spacetime 
associated with the near-horizon geometry in (\ref{MetricsType})
of our rotating RBH (\ref{RRBHTD}) in
the non-extremal case.} 
\label{vvis1}
\end{figure}

First,
$r$ is originally
restricted to a neighborhood of $r_+$
as can be seen in (\ref{seart}).
We extend its upper bound to infinity, 
while leaving its lower bound unchanged.
Next, we introduce $r^*$ 
defined in analogy with
the tortoise coordinate as follows:
\begin{eqnarray}\label{posa}
r^* = \int dr \tf^{-1}, 
\end{eqnarray}
where $\tf$ is defined in (\ref{MetricsType}). 
Then, 
we construct the maximally extended spacetime~(MES)
for the non-extremal case
by introducing the following Kruskal coordinates:
\begin{eqnarray}\label{fnben}
\begin{array}{ll}
(U_+,U_-) \equiv 
(+{e^{+\tk u_+}}/{\tk}, \,-{e^{-\tk u_-}}/{\tk}) \quad\!\!\!\!
\textrm{in the region I,} \\*[1.5mm]
(U_+,U_-) \equiv 
(+{e^{+\tk u_+}}/{\tk},\, +{e^{-\tk u_-}}/{\tk}) \quad\!\!\!\! 
\textrm{in the region II,} \\[1.5mm]
(U_+,U_-) \equiv 
(-{e^{+\tk u_+}}/{\tk},\, -{e^{-\tk u_-}}/{\tk}) \quad\!\!\!\!
\textrm{in the region III,} \\*[1.5mm]
(U_+,U_-) \equiv 
(-{e^{+\tk u_+}}/{\tk}, \,+{e^{-\tk u_-}}/{\tk}) \quad\!\!\!\! 
\textrm{in the region IV.} 
\end{array}
\end{eqnarray}
where $u_\pm \equiv t \pm r^\ast$ and $\tk$ is given in (\ref{thyrthrt}).
The MES spanned by the $(U_+,U_-)$ coordinates is shown in Fig.\,\ref{vvis1}.
Since the region between $r=0$ and the vicinity of $r_-$ 
lies outside the domain extended above, 
this region is omitted in  Fig.\,\ref{vvis1}.

On the MES,
$ds^2=\vert\tf\vert(-dt+(dr^*)^2)$ 
is given as
\begin{align}\label{eq:metric in Kruskal coordinate}
ds^2
=-\Omega ^2 \, dU_+dU_-,\quad
\Omega \equiv \vert\tf\vert^{1/2} \, e^{ -\tk  r^* }.
\end{align}
Accordingly, 
the geodesic distance between two points 
on that is given as 
\begin{align}\label{eweb}
\Omega ( x_1 ) \Omega ( x_2 ) ( U( x_2 ) -U( x_1 ) ) (V( x_1 ) -V( x_2 ) ) 
\equiv \mathbf{d}( x_1,x_2 ).
\end{align}

\section{Evaluation of the EE of the Hawking radiation}
\label{ttjuj}

In this section,
we obtain the expression of the EE
expanded in $\alpha$, 
which corresponds to the near-extremal regime of the final stage of evaporation.
Specifically, we obtain the general expression of the EE in Sec.\,\ref{sdrge}
and its form in the final stage of evaporation in Sec.\,\ref{wryus}.
Then, based on this expression, 
we solve the extremal conditions
with respect to the variables
specifying the positions of the QESs,
and determine the ultimately realized QESs
in Secs.\,\ref{wryus2} and \ref{mbst}.
Using these results, 
we obtain the expression of the EE
expanded in $\alpha$
in Sec.\,\ref{tcytc}.

\subsection{The general expression of the EE}
\label{sdrge}

In this subsection, 
we give the general expression of the EE of the HR
in the presence of the island.
\newline

\begin{figure}[H] 
\vspace{-8mm} 
\begin{center}
\hspace{-43mm} 
\includegraphics[clip,width=6.0cm,angle=-90]{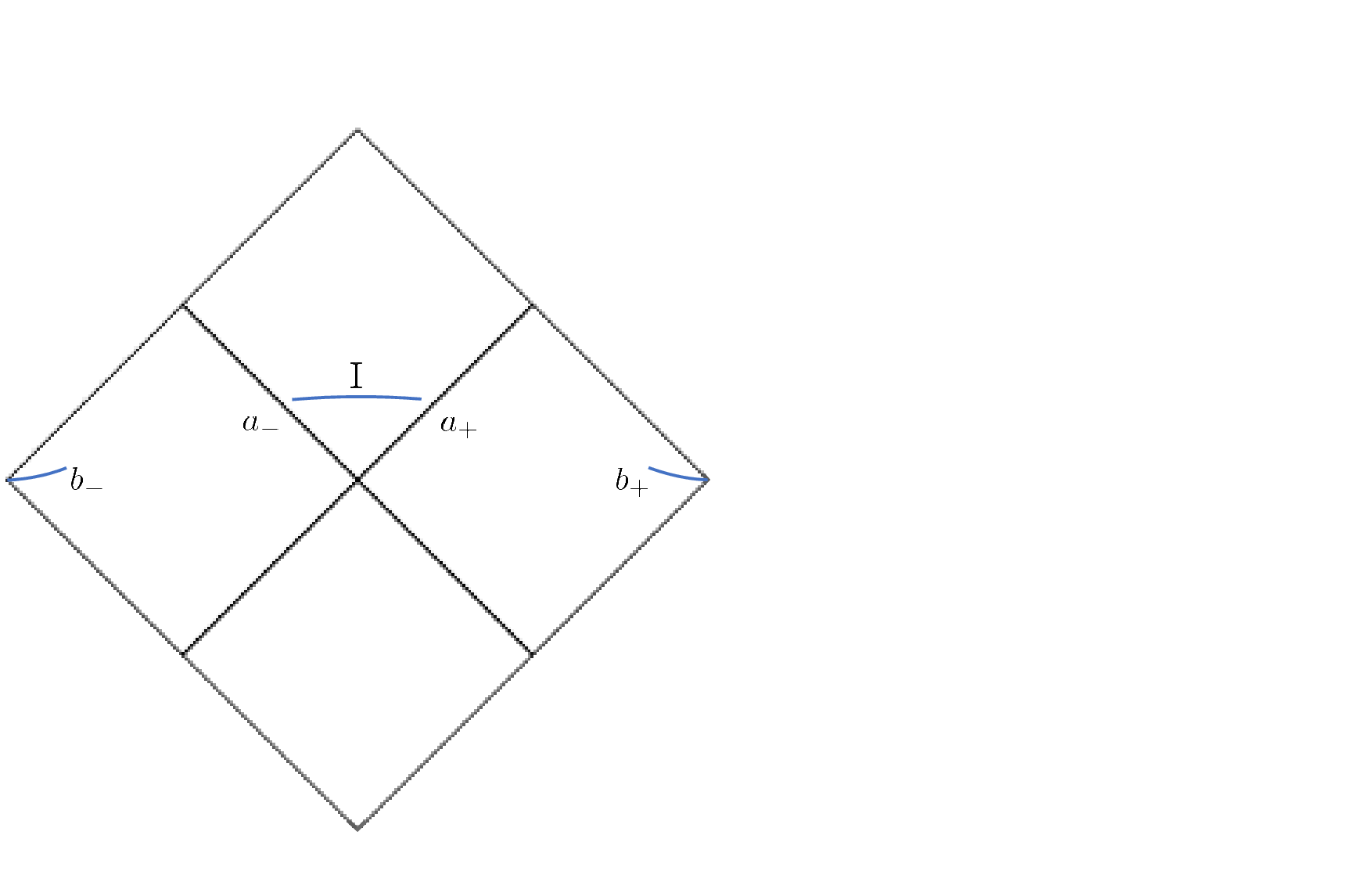} 
\end{center}
\vspace{-6.0mm}
\caption{
The blue lines 
represent 
the complementary surfaces
of the non-vanishing extremal surfaces
in the final stage of evaporation 
in the left 
figure in Fig.\,\ref{vvis1}. 
``I'' means the island.
Therefore, $a_\pm$ 
represent the positions of the tips of the non-vanishing extremal surfaces, 
and $b_\pm$ represent the positions of the cutoff surface.
According to the result of (\ref{uygu}), 
$a_+$ and $b_+$ are depicted inside $r_+$.
} 
\label{gutj}
\end{figure}

We present the position of the non-vanishing extremal surfaces
in the final stage of evaporation using $a_\pm$ and $b_\pm$ 
in Fig.\,\ref{gutj}. 
Then, we denote their coordinates as follows: 
\begin{eqnarray}\label{ntbi}
a_+:~(t_a, r_a), \quad
b_+:~(t_b, r_b), \quad
a_-:~(-t_a+i\pi/\tk, r_a), \quad
b_-:~(-t_b+i\pi/\tk, r_b),
\end{eqnarray}
and their positions in the Kruskal coordinates are obtained 
by applying these to $U_\pm$ in the region I 
in (\ref{fnben}).
Then, by following (\ref{tmrn}), 
the EE of the HR 
in the presence of the island 
is presented as follows:
\begin{eqnarray}\label{yejyk}
S_{\rm rad} = {\rm min}_X [S_{\rm island}], \quad
S_{\rm island} 
\!\! &\equiv& \!\!
2{\cal A}/4G_{\rm N} 
+ S_{\rm field}(\Sigma_{{\rm island} \,\cup\, {\rm rad}}), \quad
G_{\rm N} \sim l_p^2, 
\end{eqnarray}
where ${\cal A}$ denotes the area of a boundary $a_+$ or $a_-$, 
and the factor 2 accounts for these two boundaries.
$\Sigma_{{\rm island} \,\cup\, {\rm rad}}$ represents 
the complementary surface of the non-vanishing surface.
$S_{\rm field}(\Sigma_{{\rm rad} \,\cup\, {\rm island}})$ denotes 
the correctional contribution from the field theories. 
We use the natural units.
For more details, see Sec.\,\ref{Sec:ildfml}.
We obtain the first and second terms in (\ref{yejyk}) 
in the following.
\begin{itemize}
\item
${\cal A}$ in (\ref{yejyk}) is obtained 
using the metrics in (\ref{RRBHTD}) as
\begin{align}\label{yjyj}
{\cal A}
= 
\int_0^{2\pi} \!\! d\phi \int_0^{\pi} \! d \theta \, 
\sqrt{g_{\theta\theta}\, g_{\phi\phi}}\,\big\vert_{r=r_a} 
=& \, 
\int_0^{2\pi} \!\! d\phi \int_0^{\pi} \! d \theta \, 
(r_a^2+a^2)\sin\theta \,
\sqrt{1-(\frac{a}{r_a^2+a^2})^2\,\tD(r_a) \sin^2 \theta}
\nonumber\\*[1.5mm]
=& \,\,  
4\pi (r_a^2 + a^2)-{4a^2\pi\,\tD(r_a)}/{3(r_a^2+a^2)}+ {\cal O}(\tD(r_a)^2),
\end{align}
where $\tD$ in the non-extremal and extremal cases are defined in (\ref{wfyd}) and (\ref{tD04}), respectively.
We have taken $\tD$ up to first order, 
as $\tD$ has been taken up to leading in Sec.\,\ref{Chap:espog} 
and the present calculation is a continuation of that.
\item
$S_{\rm field}(\Sigma_{{\rm rad} \,\cup\, {\rm island}})$ 
is obtained as~\cite{Wang:2024itz}\footnote{
$
\frac
{\mathbf{d}(a_+,a_-) \mathbf{d}( b_+,b_- )}
{\mathbf{d}( a_+,b_- ) \mathbf{d}( a_-,b_+ )}
=
\Big(
\!\!\pm\!\frac
{(e^{\tk(r_b^*+t_a)}-e^{\tk(r_a^*+t_b)})(e^{\tk(r_a^*+t_a)}-e^{\tk(r_b^*+t_b)})}
{(e^{\tk((r_a^*+t_a+t_b)}+e^{\tk r_b^*})(e^{\tk r_a^*}+e^{\tk(r_b^*+t_a+t_b)})}\Big)^2
=
\Big(
\!\!\pm\!\frac
{\cosh (\tk (t_b-t_a))-\cosh (\tk(r_b^*-r_a^*))}
{\cosh (\tk (t_b+t_a))+\cosh (\tk(r_b^*-r_a^*))}\Big)^2
$. 
}\!\cite{Calabrese:2004eu,Calabrese:2005zw}
\begin{align}\label{ytvds}
S_{\rm field}(\Sigma_{{\rm rad} \,\cup\, {\rm island}}) 
=& \,\,
\frac{{\cal A}}{2\epsilon^2}+
\frac{1}{6}\ln 
\frac
{\bar{\mathbf{d}}(a_+,a_-) \bar{\mathbf{d}}( b_+,b_- ) \bar{\mathbf{d}}(a_+,b_+) \bar{\mathbf{d}}( a_-,b_- )}
{\bar{\mathbf{d}}(a_+,b_-) \bar{\mathbf{d}}( a_-,b_+ )}
\nonumber\\*[1.5mm]
=& \,\,
\frac{{\cal A}}{2\epsilon^2}
+
\frac{1}{3}\ln 
\Big\vert
\frac
{\cosh (\tk \varDelta t_{ba}^{-})-\cosh (\tk \varDelta_{ba}^*)}
{\cosh (\tk \varDelta t_{ba}^{+}) +\cosh (\tk \varDelta_{ba}^*)} 
\Big\vert 
\nonumber\\*[1.5mm]
& \hspace{+7.6mm}
+
\frac{1}{6}\ln 
\Big[
\Big(\frac{2}{\epsilon\tk}\Big)^4 \,
\vert \tf( r_a ) \vert \,
\vert \tf( r_b ) \vert
\cosh ^2( \tk t_a ) \cosh ^2( \tk t_b )
\Big], 
\end{align}
\begin{align}
& 
\hspace{-17.5mm}
\bar{\mathbf{d}}(x_1,x_2) \equiv \mathbf{d}(x_1,x_2)/\epsilon^2, 
\quad
\epsilon \sim 
\ell_p \sim 
\sqrt{\hbar G_{\rm N}/c^3} 
\approx 1.62 \times 10^{-35} \,{\rm m}, 
\nonumber
\\*[1.5mm]
& 
\hspace{-17.5mm}
\varDelta t_{ba}^{\pm} \equiv t_b \pm t_a,
\quad 
\varDelta_{ba}^* \equiv r_b^*-r_a^*
\nonumber
\end{align}
where $\mathbf{d}(x_1,x_2)$ is defined in (\ref{eweb}) 
and $\epsilon$ denotes the lattice spacing of the discretized spacetime.
$\tf$ is defined in (\ref{MetricsType}). 
\end{itemize}

Using (\ref{yjyj}) and (\ref{ytvds}),
the expression of $S_{\rm rad}$ in (\ref{yejyk}) 
is given as follows:
\begin{eqnarray}\label{yejyk1}
S_{\rm rad} 
= 
\frac{\cal A}{2G_{\rm N}} 
+
\frac{1}{3}\ln 
\Big\vert
\frac
{\cosh (\tk \varDelta t_{ba}^{-})-\cosh (\tk \varDelta_{ba}^*)}
{\cosh (\tk \varDelta t_{ba}^{+}) +\cosh (\tk \varDelta_{ba}^*)} 
\Big\vert
+
\frac{1}{6}\ln 
\big[\Big(\frac{2}{\epsilon\tk}\Big)^4 
\vert \tf( r_a ) \vert
\vert \tf( r_b ) \vert
\cosh ^2( \tk t_a ) \cosh ^2( \tk t_b )
\big],
\end{eqnarray}
where $\tf(r)$ is defined in (\ref{MetricsType}).
$1/G_{\rm N} + 1/\epsilon^2$ has been redefined as $1/G_{\rm N}$.

\subsection{$S_{\textrm{rad}}$ in the final stage of evaporation}
\label{wryus}

In this subsection, 
we obtain the expression of $S_{\textrm{rad}}$ 
in the final stage of evaporation
by taking $t_a, t_b \gg 1$. 
\newline 

The BH is in the near-extremal state for such $t_a, t_b \gg 1$.
In order to obtain $S_{\rm rad}$ at this time,
we first obtain the expression of $r^*$ from (\ref{posa}) as follows:
\begin{eqnarray}\label{resgr}
&&
r^* 
=\left\{
\begin{array}{l}
\displaystyle
\frac{2 a^2}{a-r}+2 a \ln [r-a]+r
+\Big[
\frac{2}{a^2}\Big\{
a (\frac{1}{r}-\frac{3 a}{(a-r)^2})+5 \ln [r-a]-4 \ln [r]
\Big\}
+\frac{\sqrt{\alpha}}{a^3(a-r)r}
\Big\{
\\[2.0mm] 
\displaystyle
2 a (a-11 r)+4 r (r-a) (3 \ln [r]-\ln [r-a])
\Big\}
+{\cal O}(\alpha) 
\Big]\ell_p^3
+\,C_{\rm ne}\vert_{\alpha \to 0}+{\cal O}(\ell_p^4) 
\equiv r^{*(\rm{non-ext})},
\\[6.0mm] 
\displaystyle
2r
-\frac{a^2+r^2}{r-r_{\rm ext}}+2r_{\rm ext}\ln[r-r_{\rm ext}]
+\frac{2\,\ell_p^3}{r_{\rm ext}^3}\Big\{
\frac{a r_{\rm ext}}{r}
+\frac{r_{\rm ext} (a+2 r_{\rm ext}) (a^2+r_{\rm ext}^2)}{a^2 (r-r_{\rm ext})}
\\[4.0mm] 
\displaystyle
-2 (a+r_{\rm ext}) \ln [r]
+\frac{2 (a^3+a^2 r_{\rm ext}-r_{\rm ext}^3)}{a^2}\ln [r-r_{\rm ext}]
\Big\}
+C_{\rm ext}
+{\cal O}(\ell_p^4) 
\equiv r^{*(\rm{ext})},
\end{array}
\right.
\end{eqnarray}
where 
\begin{itemize}
\item
$r^{*(\rm{non-ext})}$ 
has been obtained 
by substituting $r_\pm$ in (\ref{wred1})
and (\ref{owros})
into the expression
obtained from (\ref{posa}) with (\ref{wfyd})\footnote{
We show $r^*$ before substituting $r_\pm$ in the non-extremal case:
\begin{eqnarray}\label{resgr1}
\textrm{(\ref{posa})}\vert_{\rm (\ref{wfyd})}
\!\!\!\! &=& \!\!\!\!
r
+\frac{
(a^2+r_+^2) \ln (r-r_+)
- (r_+ \leftrightarrow r_-)
}{r_+-r_-}
-\frac{2m_0 \,\ell_p^3}{a^4 r_+^2 r_-^2 (r_+-r_-)}
\big[
\big\{
r_-^2 \,(a^2+r_+^2) (a^2+2m_0 \, r_+) \ln [r-r_+]
-(r_+ \leftrightarrow r_-) \big\}
\nonumber\\ 
&&
+r^{-1}a^2(r_+-r_-)
\big\{\!
-a^2 \, r_+ r_- 
+
\big(
2 m_0 \,r_+ r_- 
+a^2 (r_++r_-)
\big)\,r \ln [r] 
\big\}
\big]
+\,C_{\rm ne}
+{\cal O}(\ell_p^4).
\end{eqnarray}
}, while
$r^{*(\rm{ext})}$ 
has been obtained 
by substituting $r_{\rm ext}$ in (\ref{rvsrwh}) 
into the expression
obtained from (\ref{posa}) with (\ref{tD04}).
\item
$C_{\rm ne}$ and 
$C_{\rm ext}$
are the integral constants arising from the integration of (\ref{posa}). 
By using these, the arguments of the ln can be made dimensionless.
Note that $C_{\rm ne}$ depends on $\alpha$, 
and $C_{\rm ne}\vert_{\alpha \to 0}$ presents its value at the extremal limit. 
$C_{\rm ne}$ and 
$C_{\rm ext}$ do not affect results in this study as
$r^*$ always appears 
by a difference, 
as can be seen in (\ref{yejyk1}).
Therefore, we omit them in the following descriptions.
\item
$r^{*(\rm{non-ext})}$ coincides with $r^{*(\rm{ext})}$ 
up to constant\footnote{
$r^{*(\rm{non-ext})}-r^{*(\rm{ext})}=a+3\ell_p^3/a^2$ at $\alpha \to 0$, 
and this difference can be removed 
by appropriately choosing $C_{\rm ne}$.
}
at the extremal limit 
$\alpha \to 0$ 
($\alpha$ is introduced in Sec.\,\ref{Chap:tdv}).
\end{itemize}

$r_b$ denotes the position of the cutoff surface. 
Typically, such a surface locates far from the BH.
Hence, we consider the two possibilities:~$t_a,t_b \ll r_b^*$ or $r_b^* \ll t_a,t_b$ 
($r_b^*$ is obtained by substituting $r_b$ into $r^*$ in (\ref{resgr})).
$t$ is very large 
in the final stage of evaporation, 
but remains bounded above, 
as can be seen in (\ref{iiuyy}).
Consequently, it is appropriate to consider the former case, 
and therefore, in the near-extremal regime, $r_b^*$ is very large.

At this time, from the expression of $r^{*\rm{(non-ext)}}$ in (\ref{resgr}), 
we see that $r_b^*$ can be treated as $r_b^*\approx r_b$. 
On the other hand, denoting $\alpha$ 
in the final stage of evaporation 
as $\rho \alpha$~($\rho \ll 1$),
we see  $r^{*\rm{(non-ext)}}_a \approx - \sqrt{2}a/\sqrt{\gamma}$.
We assume $r_b$ in the final stage of evaporation as $r_b \gg \sqrt{\gamma}^{-1}$~(we 
will justify this assumption in the next subsection).
As a result, we can approximate $\varDelta_{ba}^*$ in the final stage of evaporation as
\begin{eqnarray}\label{dsve} 
\varDelta_{ba}^* \approx r_b.
\end{eqnarray}

Next, 
$\tk$ is near the extremal point and mostly $0$ for $t \gg 1$
as shown in Sec.\,\ref{Chap:rgtr}.
At this time, as shown in (\ref{trjtyk2}) in Appendix~\ref{App:orhwl}, 
$\tk t$ is ultimately calculated as follows:
\begin{eqnarray}\label{iiuyy}
\tk t \!\! &=& \!\!
{16(m_0^3-a^3)}/{a^{5/2}}\cdot \sqrt{{5 \pi }/{3G_{\rm N}}\cdot \delta t}+{\cal O}(\delta t),
\quad 
\delta t \equiv t^*-t \ll 1,
\quad 
0 \le t \le t^*,\nonumber\\*[1.5mm]
\!\! &\Rightarrow& \!\! 
\textrm{$\tk t  \sim \sqrt{\delta t/G_{\rm N}} \ll 1$ as $t \to t^*$},
\end{eqnarray}
where $t^*$ is defined in (\ref{tory}), 
representing the upper bound of $t$ 
at which 
the evolution of the spacetime reaches the extremal point. 
From (\ref{iiuyy}), we observe that 
$t \tk \ll 1$ in the limit $t \to t^*$, 
as noted above.
Based on this, 
we perform the following approximations:
\begin{eqnarray}\label{iiee}
\cosh (\tk (t_b \pm t_a))\approx 1, \,\,\,
\cosh (\tk t_a)\approx 1, \,\,\,
\cosh (\tk t_b)\approx 1.
\end{eqnarray}

Using (\ref{dsve}) and (\ref{iiee}),
$S_{\textrm{rad}}$ in (\ref{yejyk1}) 
can be expressed 
for $t_a$, $t_b \gg 1$ as
\begin{align}\label{yethte}
S_{\textrm{rad}}\big\vert_{t_a,t_b \gg 1}
\approx \,
\frac{\cal A}{2G_{\rm N}} 
+
\frac{1}{3}\ln 
\Big\vert
\frac{1-\cosh (\tk r_b)}{1+\cosh (\tk r_b)} 
\Big\vert
+
\frac{1}{6}\ln 
\big[
\Big(\frac{2}{\epsilon\tk}\Big)^4 \,
\vert\tf( r_a )\vert \, \vert\tf( r_b )\vert \,
\big]
\equiv S_{\textrm{rad}}^{(\textrm{late})},
\end{align}
where the second term behaves as follows:
\begin{align}\label{ktre}
\ln 
\Big\vert
\frac{1-\cosh (\tk r_b)}{1+\cosh (\tk r_b)} 
\Big\vert \approx
\left\{
\begin{array}{ll}
2\ln[\tk r_b/2]
& \textrm{for $\tk r_b \ll 1$},\\*[2.0mm]
0
& \textrm{for $\tk r_b \gg 1$}.
\end{array}
\right.
\end{align}
While $r_b$ is generally large,
$\tk$ approaches $0$ for $t_a$, $t_b \gg 1$ as can be seen in (\ref{eusrtr}).
Therefore, since the behavior of $\tk r_b$ in (\ref{ktre}) is unclear, 
we leave the second term as it is
for now.
The values of $r_a$ and $r_b$ appearing in
$S_{\textrm{rad}}\vert_{t_a,t_b \gg 1}$ 
will be determined in the following subsections. 

For comparison of our EE in (\ref{yethte}) 
with previous studies~\cite{Hashimoto:2020cas,Wang:2021woy,Kim:2021gzd,
Ahn:2021chg,Yu:2021rfg,Luongo:2023jyz,Wang:2024itz,Yu:2025euq}, 
we note that 
these studies treat $\varDelta_{ba}^*$, discussed in (\ref{dsve}), 
using a different approximation 
and the quantity $\tk t$, 
discussed in (\ref{iiuyy}), 
remains as it is. 
Because of these differences, 
our EE differs from those in the previous studies.

Typically, 
the surface gravity diverges 
in the final stage of evaporation 
for non-rotating BHs~\cite{Hashimoto:2020cas,Wang:2021woy,Kim:2021gzd,Ahn:2021chg,Yu:2021rfg,Luongo:2023jyz}, 
whereas it approaches zero 
for rotating BHs~\cite{Wang:2024itz,Yu:2025euq}, 
as shown in Sec.~\ref{Chap:rgtr}. 
Therefore, a careful analysis is required to determine 
how $\tk t$ should be treated 
in the case of rotating BHs.
In the present study, 
we have performed such an analysis, 
as presented in(\ref{iiuyy}).

\subsection{Treatment of $r_b$}
\label{wryus2}

In this subsection, 
we determine $r_b$. 
In general, 
$r_b$ is a variable chosen by hand. 
Consequently,
$r_b$ remains in final results 
as an undetermined variable .
However, since the contributions 
from $r_b$ are usually negligible or are not taken into account
at the order analyses address,
effectively no ambiguities due to $r_b$ remain in final results. 
In fact, 
this is the case 
in studies 
that analyze the Page curve 
in various BH spacetimes~\cite{Hashimoto:2020cas,
Wang:2021woy,Kim:2021gzd,Ahn:2021chg,Yu:2021rfg,
Luongo:2023jyz,Wang:2024itz,Yu:2025euq
}.
This situation is the same in the present study 
as can be seen in Sec.\,.\ref{yrtsh1}.
\newline

$\tk$ approaches $0$ 
in the limit $\alpha \to 0$ 
as can be seen in (\ref{eusrtr}).
For such behavior of $\tk$, 
we consider the following three possibilities:
\begin{align}\label{ehyut}
{\bf 1)} \,\,\, \tk r_b \ll 1, \quad \!
{\bf 2)} \,\,\, \textrm{$\tk r_b=$ a finite value}, \quad \!
{\bf 3)} \,\,\, \tk r_b \gg 1.
\end{align}
We set $r_b$ in what follows 
by exploiting the freedom in its choice.
In the previous studies~\cite{Hashimoto:2020cas,
Wang:2021woy,Kim:2021gzd,Ahn:2021chg,Yu:2021rfg,
Luongo:2023jyz,Wang:2024itz,Yu:2025euq}, 
$r_b$ is left undetermined.

\begin{itemize}
\item[-]
The cutoff surface is generally assumed to be located far away, 
which means $r_b$ is very large. 
However, $r_b$ in the case {\bf 3)} 
would be considered unphysical as it is excessively large;
therefore, we discard the case {\bf 3)}.
\item[-]
If we consider the case  {\bf 1)}, 
$\ln[\tk r_b/2]$ in (\ref{ktre}) diverges, which is undesirable. 
Therefore, making use of the fact that $r_b$ is a parameter chosen by hand,
we exclude the case {\bf 1)}.
\item[-]
The case {\bf 2)} does not suffer from the problems 
encountered in the cases {\bf 1)} and {\bf 3)},
while preserving the assumption that $r_b$ is very large. 
Therefore, we adopt the case {\bf 2)}. 
\end{itemize}

Specifically, 
we choose $r_b$ such that
$\tk r_b$ approaches a finite value 
as follows: 
\begin{align}\label{nkre}
\tk r_b \to c_{\tk r_b} \quad\!\!\!  \textrm{as ~$\alpha \to 0$},
\end{align}
where $c_{\tk r_b}$ is a finite number.
In fact, it is possible to set $r_b$ 
so that case {\bf 2)} is maintained.
Actually, $r_b$ 
obtained by solving $\partial S_{\textrm{rad}}^{( \textrm{late})}/ \partial r_b =0$
satisfies (\ref{nkre}) for $c_{\tk r_b}=\sqrt{6}$, 
which we show in (\ref{rrrst}) and (\ref{fyur}) in Appendix\,\ref{ryuyj}.
We assume that 
$c_{\tk r_b}$ is large 
so that (\ref{dsve}) holds.
This is possible 
by rescaling $r_b$ in (\ref{rrrst}) 
by an arbitrary constant, 
taking advantage of the freedom 
in the choice of $r_b$.

Substituting (\ref{nkre}) into (\ref{yethte}), 
we obtain the expression of $S_{\textrm{rad}}^{(\textrm{late})}$ as follows:
\begin{eqnarray}\label{yfete}
&& 
S_{\textrm{rad}}^{(\textrm{late})}
=
\frac{\cal A}{2G_{\rm N}} 
+
\frac{1}{6} \ln 
\big\vert
(\frac{2}{\epsilon\tk})^4 \,
\tf( r_a ) 
\big\vert
+C_{\tk r_b}, \quad
C_{\tk r_b} \equiv 
\frac{1}{3}
\ln 
\Big\vert
\frac{1-\cosh (\tk r_b)}{1+\cosh (\tk r_b)} 
\Big\vert
\bigg\vert_{\textrm{$r_b$ in (\ref{nkre})}},
\end{eqnarray}
where $r_b$ is some large number; 
therefore, $\tf(r_b) \approx 1$~($\tf(r)$ is defined 
in (\ref{MetricsType})). 
$C_{\tk r_b}$ is a function of $\alpha$, 
which approaches a finite value at $\alpha \to 0$.

For comparison between our EE in (\ref{yfete}) 
and that obtained in the previous studies~\cite{Hashimoto:2020cas,Wang:2021woy,Kim:2021gzd,
Ahn:2021chg,Yu:2021rfg,Luongo:2023jyz,Wang:2024itz,Yu:2025euq},
we have adopted the assumption (\ref{nkre}), 
whereas the previous studies do not.
This is a difference 
between the present study and the previous studies.
Therefore, 
together with the differences
associated with (\ref{dsve}) and (\ref{iiuyy}),
our EE in (\ref{yfete})
differs from that obtained in the previous studies. 

\subsection{The extremum of $r_a$}
\label{mbst}

In this subsection, 
using $S_{\textrm{rad}}^{(\textrm{late})}$ in (\ref{yfete}), 
we solve the extremal condition 
$\partial S_{\textrm{rad}}^{( \textrm{late})}/ \partial r_a =0$
and determine the extremum $r_a$. 
\newline

$\partial S_{\textrm{rad}}^{( \textrm{late})}/ \partial r_a$
is calculated as follows:
\begin{align}\label{epowe}
\frac{\partial S_{\textrm{rad}}^{( \text{late} )}}{\partial r_a}
=
\frac{1}{2G_{\rm N}}
\frac{\partial {\cal A}}{\partial r_a}
+
\frac{1}{6}
\frac{\tf'(r_a)}{\tf(r_a)}
=& \,\,
\frac{1}{2G_{\rm N}} 
\frac{4\pi}{3} 
\Big(
6 r_a 
+
\frac{a^2}{(a^2+r_a^2)^2}
(2 r_a \Delta (r_a)-(a^2+r_a^2) \Delta '(r_a))
\Big)
\nonumber\\*[1.5mm]
& \hspace{-20mm}
-\frac{ a (a+r_a)}{3(a-r_a) ( a^2+r_a^2 )}
+\frac{\alpha(a+r_a) }{3(r_a-a)^3}
+{\cal O}(\alpha^2)
+\frac{1}{3 a^2}
\Big\{
\Big(
\frac{3}{(a-r_a)^2}
-\frac{2(a+r_a)}{r_a^3}
\Big)
\nonumber\\*[1.5mm]
& \hspace{-20mm}
+
\frac{2\alpha(-a^4+2 a^3 r_a+2 a^2 r_a^2+4 a r_a^3+2 r_a^4)}{r_a^3 (a-r_a)^4}
+{\cal O}(\alpha^2)
\Big\}\ell_p^3
+{\cal O}(\ell_p^4),
\end{align}
where $\tf'(r_a)$ and $ \Delta '(r_a)$ denote 
$\partial \tf(r_a)/ \partial r_a$ and $\partial \Delta(r_a)/ \partial r_a$, respectively. 
As for $\cal A$ and  $\Delta(r)$, 
we used the ones given  in (\ref{yjyj}) and  (\ref{nklsd}), respectively. 
In the previous studies~\cite{Hashimoto:2020cas,Wang:2021woy,
Kim:2021gzd,Ahn:2021chg,Yu:2021rfg,Luongo:2023jyz,
Wang:2024itz,Yu:2025euq}, 
the equation 
corresponding to $\textrm{(\ref{epowe})}=0$ is approximately solved 
by assuming that $r_a$ lies near $r_+$~(at this time, 
the island formula 
is implicitly used).
We also solve $\textrm{(\ref{epowe})}=0$ in this way.
Specifically, 
we replace $r_a$ in (\ref{epowe}) with $r_ + + \delta$, 
then expand it in $\delta$, which results in as follows:
\begin{align}\label{syirip}
{(\ref{epowe})}\big\vert_{r_a =r_+ + \delta}
=
\Big(
\frac{1}{3 \delta}
-\frac{1}{6 a}
+\frac{4a \pi}{G_{\rm N}}
+{\cal O}(\delta)
\Big)
+
\Big(
\frac{4}{3 a^4}(\frac{9 \pi  a^2}{G_{\rm N}}-1)
+{\cal O}(\delta)
\Big)\ell_p^3
+{\cal O}(\ell_p^4)
+{\cal O}(\sqrt{\alpha})
=0.
\end{align}
By solving this equation for $\delta$,  
we obtain the extremum of $r_a$ as follows: 
\begin{align}\label{uygu}
r_a
=r_++\delta
=r_+ +(-{1}/{12a\pi}+{\ell_p^3}/{4a^4\pi}+{\cal O}(\ell_p^4))G_{\rm N}
+{\cal O}(G_{\rm N}^2)
\approx r_+  -{G_{\rm N}}/{12a\pi},
\end{align}
where $G_{\rm N} \sim \ell_p^2$ as can be seen in (\ref{ytvds}), 
and the present analysis is up to $\ell_p^3$.

Typically $r_a$ is located 
outside of the horizon,~i.e.~$r_+ < r_a$~\cite{Hashimoto:2020cas,
Wang:2021woy,Kim:2021gzd,Ahn:2021chg,Yu:2021rfg,Luongo:2023jyz,
Wang:2024itz,Yu:2025euq}~\cite{Almheiri:2019yqk,He:2021mst}.
In contrast, our result lies in the inside.
This is presumably because
the second term of (\ref{yfete}) 
becomes independent of $r_a$
as a result of the approximation in (\ref{dsve}).

\subsection{The EE of the HR in the final stage of evaporation}
\label{tcytc}

$r_a$ has been  determined in (\ref{uygu}),  
and $r_b$ has been chosen as in  (\ref{nkre}).
Using these in (\ref{yfete}), 
the EE of the HR 
at the near-extremal regime 
in the final stage of evaporation 
is obtained as follows:
\begin{eqnarray}\label{ehtjf}
S_{\textrm{rad}}^{(\textrm{late})} \big\vert_{\textrm{(\ref{nkre}), (\ref{uygu})}}
\!\! &=& \,\,\,\,\hspace{0.5mm}
\Big\{
\Big(
\frac{4 \pi  a^2}{G_{\rm N}}
-\frac{1}{3}
+{\cal O}(G_{\rm N})
\Big)
+
\Big(
-\frac{1}{6 a^3}
+{\cal O}(G_{\rm N})
\Big)
\ell_p^3
+{\cal O}(\ell_p^4)
\Big\}
\nonumber\\*[1.5mm]
&&
+ \,
\Big\{
\Big(
\frac{8 a^{3/2} \pi}{G_{\rm N}}
-\frac{5}{9 \sqrt{a}}
+{\cal O}(G_{\rm N})
\Big)
+
\Big(
\frac{4 \pi }{a^{3/2} G_{\rm N}}
+\frac{4}{3 a^{7/2}}
+{\cal O}(G_{\rm N})
\Big)
\ell_p^3
+{\cal O}(\ell_p^4)
\Big\}
\sqrt{\alpha}
+{\cal O}(\alpha)
\nonumber\\*[1.5mm]
&&
+ \,
\Big\{
\Big(
\frac{1}{6} \ln [\frac{2 a^2 G_{\rm N}^2}{9 \pi ^2 \alpha ^2 \epsilon ^4}]
+{\cal O}(G_{\rm N})
\Big)
+
\Big(
\frac{12 \pi }{a G_{\rm N}}+\frac{1}{3 a^3}
+{\cal O}(G_{\rm N})
\Big)
\ell_p^3
+{\cal O}(\ell_p^4)
\Big\}
\nonumber\\*[1.5mm]
&&
+ \,
\Big\{
\Big(
-\frac{8 \left(\pi  a^{3/2}\right)}{G_{\rm N}}
+\frac{2 \sqrt{2}-1}{3 \sqrt{a}}
+{\cal O}(G_{\rm N})
\Big)
+ 
\Big(
\frac{288 \pi ^2 \sqrt{a}}{G_{\rm N}^2}+\frac{4 \sqrt{2}-{8}/{3}}{a^{7/2}}
+{\cal O}(G_{\rm N})
\Big)
\ell_p^3
\nonumber\\*[1.5mm]
&&
\hspace{72.5mm}
+\,{\cal O}(\ell_p^4)
\Big\}
\sqrt{\alpha}
+{\cal O}(\alpha)
+C_{\tk r_b},
\nonumber\\*[1.5mm]
&\equiv& \!\! \Lambda_1+\Lambda_2,
\end{eqnarray}
where 
\begin{itemize}
\item
we have obtained the result above 
by calculating 
using $\cal A$ and $\Delta(r)$  given in (\ref{yjyj}) and (\ref{nklsd}), respectively, 
as well as (\ref{epowe}).
\item
The first and second lines of the above expression originate from the first term in (\ref{yfete}), 
whereas the third and fourth lines originate from the second term in (\ref{yfete}). 
We have defined these contributions as $\Lambda_1$ and $\Lambda_2$, respectively, as follows:
\begin{eqnarray}\label{wyeec}
\Lambda_1 \equiv
\textrm{the first and second lines in (\ref{ehtjf})}, \quad
\Lambda_2 \equiv
\textrm{the third and forth lines in (\ref{ehtjf})}.
\end{eqnarray} 
\item
Since this study focuses on the small $\alpha$ regime, 
we first performed the expansion of $\alpha$ in the above expression.
If the expansions of $l_p$ and $G_N$ are firstly performed,
the perturbative expression breaks down 
in the limit $\alpha \to 0$, 
which we show 
in the footnote\footnote{
We show 
$S_{\textrm{rad}}^{(\textrm{late})} \big\vert_{\textrm{(\ref{nkre}), (\ref{uygu})}}$
expanded in $l_p$ firstly, then $G_N$ secondly, and $\alpha$ lastly in the following: 
\begin{align}\label{vtwew}
S_{\textrm{rad}}^{(\textrm{late})} \big\vert_{\textrm{(\ref{nkre}), (\ref{uygu})}} 
=& \,\,\,\,\,\,\,
\big\{
\big(
4 a^2 \pi 
+8 \pi \sqrt{a^3\alpha}
+{\cal O}(\alpha)
\big)
/{G_{\rm N}}
+
\big(
-{1}/{3}
-{5}/{9}\cdot\sqrt{{\alpha}/{a}}
+{\cal O}(\alpha)
\big)
+{\cal O}(G_{\rm N})
\big\}
\nonumber\\*
&
+
\big\{
\big(
4\pi \sqrt{{\alpha}/{a^3}}
+{\cal O}(\alpha)
\big)
/G_{\rm N}
+
\big(
-{1}/{6a^3}
+{4}/{3}\cdot\sqrt{{\alpha}/{a^7}}
+{\cal O}(\alpha)
\big)
+{\cal O}(G_{\rm N})
\big\}
\ell_p^3
+{\cal O}(\ell_p^4)
\nonumber\\*
&
+
\big\{
\big(
{1}/{6}\cdot \ln \big\vert {64 a^5}/{\alpha \varepsilon ^4}\big\vert
+{2(-1+\sqrt{2})}/{3}\cdot\sqrt{{\alpha}/{a}}
+{\cal O}(\alpha)
\big\}
+{\cal O}(G_{\rm N})
\big\}
\nonumber\\*
&
+
\big\{
-{1}/{\sqrt{a^5\alpha}}
-{1}/{2a^3}
+{(-17+12\sqrt{2})}/{3}\cdot\sqrt{{\alpha}/{a^7}}
+{\cal O}(\alpha)
+{\cal O}(G_{\rm N})
\big\}\ell_p^3
+{\cal O}(\ell_p^4)
+C_{\tk r_b}.
\end{align}
We can see that 
the expression above breaks down
in the limit $\alpha \to 0$
due to a term in the fourth line. 
}.
\item
At the end of Secs.\,\ref{wryus}, \ref{wryus2} and \ref{mbst},
we have commented on 
the comparison with previous studies 
and the assumptions used to obtain this result.
\end{itemize}

\section{The analysis on the emergence of the remnant}
\label{yrtsh}

In (\ref{ehtjf}), 
we have obtained the expression of the EE of the HR 
considered in this study, from which we can make the following observations:
\begin{itemize}
\item[1.)] 
In general, the qualitative behavior of a quantity 
expressed perturbatively 
is determined
by its leading term,
and the leading term of $\Lambda_1$ is ${4\pi a^2}/{G_{\rm N}}$.
From this, we observe that 
$\Lambda_1$ is positive for all $\alpha \ge 0$ 
and approaches ${4\pi a^2}/{G_{\rm N}}$ 
from above as $\alpha \to 0$.

The mass in the present study is presented as $m=m_{\rm ext}+\alpha$.
Therefore, the limit $\alpha \to 0$ corresponds to the extremal limit, 
whereas large values of $\alpha$ correspond to
the early stage of the evaporation process.

In the present study, $a$ is assumed 
to remain a finite constant 
throughout the evaporation process, 
as discussed in Sec.\,\ref{Chap:rgtr}.
Therefore, the BH remains of finite size 
when it reaches the extremal limit, and the term 
${4\pi a^2}/{G_{\rm N}}$ survives 
in the limit $\alpha \to 0$.

\item[2.)] 
Roughly speaking, $\Lambda_2$ behaves as 
$
\ln [\frac{2 a^2 G_{\rm N}^2}{9 \pi ^2 \alpha ^2 \epsilon ^4}] 
\sim -2\ln [\alpha]+\ln [\frac{2 a^2}{9 \pi ^2}] 
$, 
where we have used the relation $G_{\rm N} \sim \ell_p^2 \sim \epsilon^2$ 
noted in (\ref{ytvds}). 
$a$ 
is assumed to remain a finite constant throughout this study.
Therefore, $\Lambda_2$ decreases logarithmically as $\alpha$ increases.
On the other hand, as $\alpha$ decreases, 
$\Lambda_2$ crosses zero from below at a certain value of $\alpha$, 
and thereafter becomes positive and increases logarithmically. 
We denote the value of $\alpha$ at which $\Lambda_2$ crosses zero by $\alpha_2$.
\item[3.)]  
Page curves usually start and end at $S_{\rm rad}=0$. 
However, as discussed in Sec.\,\ref{Chap:rgtr}, 
the rotation parameter $a$ 
in the present study is assumed 
to be a finite constant
throughout the evaporation process. 
Therefore, in the present study, 
the BH remains finite in size 
when it reaches the extremal limit 
Accordingly, the Page curve in the present study is 
zero
only at the beginning of the evaporation process, 
whereas it remains finite at late times.
Therefore, the condition 
$\blue{\S4852}=0$ is expected to admit only one large-$\alpha$ solution.
We denote this $\alpha$ by $\alpha_1$.
\end{itemize} 
Then, based on the general property that the entropy is always non-negative\footnote{
Suppose we have a density matrix. 
In the case of a pure state, 
it is given by
$\rho=\vert \Psi\rangle \langle \Psi\vert$.
For a mixed state with
Hamiltonian $H$ at temperature $\beta^{-1}$, 
$\rho$ is denoted as $e^{-\beta H}/Z$, 
where $Z=\sum e^{-\beta H}$.
In both cases, we can see that $\rho$ is positive semi-definite, 
and all its eigenvalues lie in the interval $[0,1]$.
Using this $\rho$, 
the von Neumann entropy $S$ is given by $- \Tr [\rho \ln \rho]$.
Hence, $0 \le S$~(this can also be seen 
from the original definition 
$S\equiv k_{\rm B} \ln W$~($W$ denotes number of states)).

Next, 
we divide the entire system into two subsystems $A$ and $B$.
At this time, the EE between them is given by
$- \Tr [\rho_{\rm A} \ln \rho_{\rm A}] \equiv S_{\rm A}$, 
where $\rho_{\rm A} \equiv \Tr_{\rm B} [\rho]$
is the reduced density matrix of subsystem $A$~($\Tr_{\rm B}$ denotes the trace 
over the degrees of freedom in $B$).
Since $\rho$ is the density matrix introduced above, 
$\rho_{\rm A}$ is also positive semi-definite, 
and all its eigenvalues lie in the interval $[0,1]$. 
Hence, $0 \le S_{\rm A}$.
},
we find that $\alpha_1$ in 3.) and $\alpha_2$ in 2.) are 
the upper and lower bounds for $\alpha$, respectively,
as explained below:
\begin{itemize}
\item[i.)] 
$\S4852$ must be non-negative, and
$\alpha_1$ corresponds to the beginning of the evaporation process.
Therefore, $\alpha_1$ is the upper bound on $\alpha$, 
and the allowed values of $\alpha$ satisfy $\alpha \le \alpha_1$.

$\alpha_1$ is expected to be very large. 
By contrast, the present analysis is applicable 
only in the small-$\alpha$ regime.
Indeed, it should be noted that  
$\alpha$ is assumed throughout this study to be small 
compared with $m_{\rm ext} \sim a$, 
as can be seen from (\ref{wred}). 
In fact, 
(\ref{ehtjf}) is a perturbative expression valid 
only for $\alpha \ll a$~(which we refer to as the small-$\alpha$ regime).
Accordingly, the present discussion is restricted 
to the small-$\alpha$ regime. 

Therefore, 
although it is possible to show that 
$\alpha_1$ does not \blue{lie in} the small-$\alpha$ regime~(as shown in the next subsection) 
and thereby infer that $\alpha_1$ is very large,
the value of $\alpha_1$ cannot be determined 
within the framework of the present study.
To establish the point above,
it must be also shown that
$\S4852$ is positive in the small-$\alpha$ regime,
if we argue that $\alpha_1$ is an upper bound for $\alpha$.

\item[ii.)]  
$\Lambda_1$ and $\Lambda_2$ in (\ref{ehtjf})
are the contributions 
from the bulk gravity and the bulk field theory,
respectively, as mentioned 
in the third point after (\ref{veri}).
Therefore,
we assume that
their signs remain unchanged,
respectively,
throughout the evaporation process.

Under this assumption,
$\Lambda_2$ is expected to remain negative
throughout the evaporation process.
Hence, $\alpha_2$, 
the point at which $\Lambda_2$ changes sign,
is identified  as the lower bound on $\alpha$.
\end{itemize}
In the following subsections, 
we show that $\alpha_1$ does not lie 
in the small-$\alpha$ regime, 
determine $\alpha_2$,
and then discuss their physical implications.

\subsection{The upper bound on $\alpha$} 
\label{yrtsh1}

In this subsection, 
we solve (\ref{ehtjf}) 
and show that $\alpha_1$ does not lie 
in the small-$\alpha$ regime. 
In addition, we note that 
$\S4852$
is positive in the small-$\alpha$ regime.
\newline

As mentioned in i.), $\alpha_1$
denotes the upper bound of $\alpha$, 
which is obtained 
by solving
$S^{\rm (late)}_{\rm rad}=0$. 
However, it cannot be solved 
if $S^{\rm (late)}_{\rm rad}$ is as it is.
Therefore, we focus on its leading terms as  
\begin{align}\label{reue}
\textrm{$\S4852$ in (\ref{ehtjf})}
\approx& \,\,
\underbrace{
\frac{4 a^2\pi }{G_{\rm N}}
-\frac{1}{6a^3}\ell_p^3
}_{\equiv \,\widetilde{\Lambda}_1} 
+
\underbrace{
\frac{1}{6} \ln
\blue{\Big[
\frac{2 a^2 G_{\rm N}^2}{9 \pi ^2 \alpha ^2 \epsilon ^4}
\Big]}
+
\frac{12\pi}{aG_{\rm N}} 
\ell_p^3,
}_{\equiv \,\widetilde{\Lambda}_2} 
\end{align}
where $\epsilon$ is introduced 
in (\ref{ytvds}) and $\ell_p$ is introduced in (\ref{RRBHTD}).
$\ell_p$ represents the effect of the regularization for the BH singularity.

Then, \blue{solving} (\ref{reue}) $=c$, 
we obtain the following $\alpha$:
\begin{eqnarray}\label{reuere1}
\alpha
\!\!\! &=& \!\!\!
\frac{\sqrt{2} a G_{\rm N}}{3 \pi  \epsilon ^2}
e^{
-3c
-{\ell_p^3}/{2 a^3}
+{12 \pi (a^3+3 \ell_p^3)}/{a G_{\rm N}}
}
\nonumber\\*[1.5mm]
\!\!\! &=& \!\!\!
e^{\frac{12 a^2 \pi }{G_{\rm N}}-3c+{\cal O}(G_{\rm N}^4)}
\Big\{
\Big(
\frac{\sqrt{2}a}{3\pi} \frac{G_{\rm N}}{\epsilon ^2}+{\cal O}(G_{\rm N}^4)
\Big)
+
\Big(
12\sqrt{2}
-\frac{G_{\rm N}}{3 \sqrt{2} \pi  a^2 }
+{\cal O}(G_{\rm N}^4)
\Big) 
\frac{\ell_p^3}{\epsilon ^2}
\Big\}
+{\cal O}(\ell_p^4),
\end{eqnarray}
where $c$ is a parameter 
introduced to denote the value of 
$\S4852$
in (\ref{reue}).
By taking the limit $c \to 0$ 
in the above expression for $\alpha$,
$\alpha_1$ is obtained as
\begin{eqnarray}\label{reuere}
\textrm{$\alpha$ in (\ref{reuere1})} 
\xrightarrow[c \to 0]{}
\alpha_1 \sim e^{{12a^2\pi}/{G_{\rm N}}}, 
\end{eqnarray}
where we have used the relations $\epsilon \sim \ell_p$ and $G_{\rm N} \sim \ell_p^2$ 
as noted in (\ref{ytvds}).

From the result above, 
we observe that $\alpha_1$ takes a very large value.
Then, as stated in i) 
at the beginning of this section, 
such large values of $\alpha$ cannot be discussed in the present study.
However, it follows from this result that 
$\alpha_1$ does not exist in the small-$\alpha$ regime. 
In addition, 
the fact that 
only one such large $\alpha_1$ is obtained 
is consistent with our expectation 
stated in 3).  

From the above conclusion, 
it follows that 
the small-$\alpha$ regime corresponds to the late stage of the evaporation.
If this is the case, $\S4852$ must be positive in the small-$\alpha$ regime. 
This can be seen immediately from (\ref{reue}) 
together with $\epsilon \sim \ell_p$ and $G_{\rm N} \sim \ell_p^2$.
 
Lastly, we make one remark.
We find that $\alpha_1 \sim 1$ when $a=0$.
This implies that
$\alpha_1 \sim 1$
in the non-rotating case,
which appears counterintuitive,
since $\alpha_1$ is expected to remain large
in the non-rotating case.
However, as discussed in Sec.\,\ref{Chap:rgtr},
the present analysis assumes that
$a$ remains a finite constant
throughout the evaporation process.
Therefore, 
the case $a=0$ lies outside the scope of the present analysis.

\subsection{The lower bound on $\alpha$} 
\label{yrtsh2}

We first describe the solution for $\alpha$ 
when $\widetilde{\Lambda}_2=c$ as follows:
\begin{align}\label{retmh}
\alpha
=& \,\,
\frac{\sqrt{2}}{3\pi} \,a
\,e^{-3c
+36 \pi \, {\ell_p^3}/{a G_{\rm N}}
}\,
\frac{G_{\rm N}}{\epsilon ^2}
=
\Big(
\frac{\sqrt{2}a}{3\pi} \, e^{-3c} \, \frac{G_{\rm N}}{\epsilon ^2} 
+{\cal O}(G_{\rm N}^2)
\Big)
+12\sqrt{2} \, e^{-3c} \, \frac{\ell_p^3}{\epsilon ^2}
+{\cal O}(\ell_p^4)
\nonumber\\*[1.5mm]
\sim& \,\,
\blue{\frac{\sqrt{2}}{3\pi} \, a \, e^{-3c} + 12\sqrt{2} \, e^{-3c} \,\ell_p,}
\end{align}
where $c$ is simply a parameter 
introduced to denote 
the value of $\widetilde{\Lambda}_2$, 
as well as (\ref{reuere1}). 
We have used the relations 
$\epsilon \sim \ell_p$ and $G_{\rm N} \sim \ell_p^2$ 
\blue{noted} in (\ref{ytvds}). 

From the result above, 
we observe that $\alpha$ decreases 
as $c$ increases, and
$\widetilde{\Lambda}_2=c<0$ is expected to hold
throughout the evaporation process, 
as mentioned in ii.) at the beginning of this section.
Then,
large values of $\alpha$ correspond to the early stage of the evaporation process 
as discussed in the previous subsection, 
and $\alpha$ continues to decrease from its large initial value 
as the evaporation proceeds.
Therefore, 
the value of $\alpha$ at which $\widetilde{\Lambda}_2$ crosses zero
is identified as the lower bound on $\alpha$.
In 2), we have denoted the lower bound on $\alpha$ as $\alpha_2$.
Hence, $\alpha_2$ is obtained 
by setting $c = 0$ in (\ref{retmh}) as
\begin{align}\label{ehgyt}
\alpha_2 
\sim {\sqrt{2}\,a}/{3\pi} + 12\sqrt{2} \,\ell_p
\sim {\sqrt{2}\,a}/{3\pi}.
\end{align}

It is then assumed that 
the evaporation of the BH terminates 
at $m=m_{\rm ext}+\alpha_2$.
Therefore,
the nonzero finite value of $\alpha_2$ 
obtained in (\ref{retmh})
implies that 
the evaporation terminates before reaching the extremal limit, 
and hence a remnant eventually forms in the rotating RBH.
This constitutes the main result of the present study.

Here, 
\begin{itemize}
\item
the coefficient ${\sqrt{2}}/{3\pi}$
makes the value of $\alpha_2$ smaller 
compared with $m_{\rm ext}$.
As mentioned after (\ref{reuere}), 
in this study, 
$\alpha$ is assumed to be small 
compared with $m_{\rm ext} \sim a$, 
and the result (\ref{ehgyt}) does not contradict this assumption 
due to the factor ${\sqrt{2}}/{3\pi}$.
\item
It can be seen that
$\alpha_2$ vanishes when $a=0$.
However, the case $a=0$ lies 
outside the scope of the present analysis. 
The reason is explained after (\ref{reuere}).
\item
As mentioned above, 
although the values of $G_{\rm N}$ and $\epsilon$ are comparable to each other,
we have not determined their exact values.
Therefore,
${G_{\rm N}}/{\epsilon ^2}$ 
may be of order unity, larger than 1, or smaller than 1.
However, determining which case applies 
is beyond the scope of the present study. 
Therefore, in (\ref{retmh}), we 
proceeded under the assumption that 
${G_{\rm N}}/{\epsilon ^2}\sim 1$.
\item
Regarding the effect of $\ell_p$, 
which was specifically introduced in this study, 
even if we take the limit $\ell_p \to 0$  in (\ref{ehgyt}), 
the result in (\ref{ehgyt}) is not changed essentially.
Therefore, it is just a correction
in the emergence of the remnant.
\end{itemize}

\section{Summary}
\label{wh54h} 

In this study, 
from an analysis of the entanglement entropy~(EE) 
in the final stage of evaporation, 
we obtained an indication that
a remnant eventually forms
in a regular rotating black hole~(BH).

The fundamental idea in our approach was
to introduce a parameter $\alpha$ as in (\ref{wred}) 
and determine its value 
at which 
the correction term in the EE 
vanishes as its lower bound, 
based on the fact that 
EE is generally non-negative. 
As a result, 
in Sec.\,\ref{yrtsh2}, 
we obtained an indication that
a remnant ultimately forms. 

As mentioned in Sec.\,\ref{lcahs},
the issue of whether 
a remnant ultimately forms 
is important 
for understanding the time evolution of BH spacetimes.
However, explicit analyses of this issue have not progressed 
significantly so far. 
Under such circumstances, 
the present study has yielded a clear result regarding this issue.

We used the island formula 
when approximating $r_a \approx r_+$
in Sec.\,\ref{mbst}, 
and
our calculation of the EE was carried out 
largely in the same manner as in recent studies 
concerning the Page curve~\cite{
Hashimoto:2020cas,Wang:2021woy,
Kim:2021gzd,Ahn:2021chg,
Yu:2021rfg,Luongo:2023jyz,
Wang:2024itz,Yu:2025euq
}. 
However, there are several differences from their analyses.
We calculated the EE 
under assumptions different 
from those adopted in these previous studies.
We have commented on this point after (\ref{ehtjf}).
In addition,
they consider a non-evaporating BH, 
whereas we considered a shrinking BH 
by introducing the parameter $\alpha$. 

In this study, 
motivated by the possibility that 
the structure of the central region of the BH 
may ultimately become important,
we considered a regular BH.
However, 
its effect was found to be only 
a correction to the final result.
Nevertheless, the equation 
determining the horizon radii 
was given by a quintic equation 
as in (\ref{RRBHTD}), 
and solving it was never straightforward.
We solved it perturbatively. 
We then obtained the horizon radii 
in the non-extremal and extremal cases and 
demonstrated that they are smoothly connected.
To the best of our knowledge, 
so far, there have been no studies 
other than the present one 
that explicitly obtain the horizon radii and demonstrate that 
they connect smoothly. 

As a future direction, 
since we have obtained an indication of remnant formation, 
it would be worthwhile to investigate the underlying mechanism.
For references discussing this mechanism, 
we refer the reader to Sec.~\ref{lcahs}.
\newline 

\noindent
{\bf Acknowledgment:~}Dr.\,Akihiro Miyata 
carefully reviewed the manuscript of this study and provided many important comments. 
These had a significant impact on the description of this study. 
The author would like to express his deepest gratitude to him.
\appendix

\section{Introduction to rotating regular rotating black hole}
\label{App:RBH} 

In this Appendix, 
we review the metrics of a regular black hole~(RBH) and introduce a rotating RBH. 
We finally comment on the fact that RBHs generally do not satisfy the field equation.
\newline

There exists a class of black hole~(BH) spacetimes 
known as the RBH spacetime,
in which the central region is regularized 
by the de Sitter space and is not the singularity\,\cite{Bardeen,Dymnikova:1992ux}.
The typical metrices for a non-rotating RBH are given by
\begin{eqnarray}\label{gfshj}
- g_{tt} = \frac{1}{g_{rr}} 
= 1-\frac{2\,\tilde{m}}{r}
\sim
\left\{
\begin{array}{lll} 
\displaystyle 1 - {2m_0}/{r} & \textrm{(Schwarzschild)} & \textrm{for $r \gg L$}\\[1.5mm]
\displaystyle 1 - {r^2}/{L^2} & \textrm{(de Sitter)} & \textrm{for $r \sim 0$}
\end{array} 
\right., 
\quad 
\tilde{m} \equiv \frac{m_0 r^3}{r^3 + m_0L^2},
\end{eqnarray}
where 
$m_0$ is the mass of the BH and 
$L$ is a small parameter with dimensions of length, 
which we take as the Plank length $\ell_p$ in this study, 
as can be seen in (\ref{RRBHTD}).
The metrices in the angular directions remain unchanged. 
``(de Sitter)'' and ``(Schwarzschild)'' indicate the type of spacetime in those regions.
The de Sitter core of the RBH is interpreted 
as an effective description of quantum gravitational effects.

Currently, rotating RBHs are constructed 
by incorporating a rotational effect 
into non-rotating ones 
using the Newman-Janis algorithm~\cite{Bambi:2013ufa}.
The rotating RBH considered in this study is also the one 
obtained via this algorithm.

The metric (\ref{gfshj}) does not satisfy the gravitational field equations.
This is because, 
the cosmological constant needs
to vary on each point on the spacetime
for this purpose.
Numerous studies have addressed this issue, 
which can be categorized into the following three types:
{\bf 1)}\, 
the type that connects the outside Schwarzschild region and the de Sitter core 
via a thin shell~\cite{Uchikata:2012zs,Uchikata:2014kwm},
{\bf 2)}\, 
the type that introduces additional effects into the model, 
such as non-linear electromagnetic fields~\cite{Ayon-Beato:1998hmi,
Ayon-Beato:1999kuh,
Bronnikov:2000vy,Ayon-Beato:2004ywd,
Hayward:2005gi,Fan:2016hvf
}, 
{\bf 3)}\, 
the type that considers non-commutative spacetime~\cite{Nicolini:2005vd,Modesto:2010rv}. 
Currently, it is possible to construct a Lagrangian that 
admits a non-rotating RBH as a solution~\cite{Fan:2016hvf}, 
and the framework of~\cite{Fan:2016hvf} has been further developed in, 
for example,~\cite{Isomura:2023oqf,Tsuda:2023tcs,Tsuda:2026xjc}. 
In contrast, no study has yet demonstrated a Lagrangian 
that admits rotating RBHs as solutions.
Since the fact that the RBH is not an exact solution 
does not affect the applicability of the generalized Ryu-Takayanagi formula, 
we proceed with this study.

\section{The horizon radii in $\tD=0$ reduced to a quartic equation}
\label{App:Delta4} 

In Sec.\,\ref{Sec:RBH}, 
we solved a quintic equation $\tD=0$ 
up to $\ell_p^3$ order.
In this Appendix, 
we solve it 
by an alternative way,
which is to approximately reduce $\tD$ to a quartic equation. 
\newline

First, we expand $\tD$ up to $\ell_p^3$ order, 
and define $\tD_4$ as follows:
\begin{eqnarray}\label{ladv}
\tD 
= \frac{r^3 (r^2- 2m_0 r+ a^2 ) + (r^2 +a^2)\ell_p^3}{r^3+\ell_p^3} 
=
\underbrace{
r^2 - 2m_0 r + a^2 + \frac{2m_0}{r^2} \, \ell_p^3}_{\equiv \tD_4} + \cdots. 
\end{eqnarray}
Then, denoting its four solutions 
in the non-extremal case as 
$r_{\rm I\pm}$ and $r_\pm$, 
these are obtained as follows:
\begin{subequations}\label{rtym}
\begin{align} 
\label{rtym2}
r_\pm 
=& \,
\frac{1}{2}
(m_0+\sqrt{
\Omega_1
}
\pm 
\sqrt{
\Omega_{2+}
}
\,) 
=\, 
m_0 \pm \eta
\mp \frac{m_0}{(m_0\pm\eta)^2\eta}\ell_p^3
+\cdots, 
\\* 
\label{rtym1}
r_{\rm I \pm} 
=& \, 
\frac{1}{2}
(m_0-
\sqrt{
\Omega_1
}
\pm 
\sqrt{
\Omega_{2-}
}
\,)=
\pm\frac{i \sqrt{2m_0} }{a}\ell_p^{3/2}
-\frac{2 m_0^2}{a^4}\ell_p^3
\pm\frac{i \sqrt{2} m_0^{3/2} \left(a^2-5 m_0^2\right)}{a^7}\ell_p^{9/2} 
+\cdots,
\end{align} 
\end{subequations}
\vspace{-4mm}
\begin{eqnarray}
\hspace{-15.0mm}
\Omega_1 \!\! &\equiv& \!\! 
\kappa+\theta,\quad
\Omega_{2\pm} \equiv 2\theta-\kappa
\pm
{2m_0 \eta^2 }/{\sqrt{\kappa+\theta}},
\nonumber\\* 
\kappa \!\! &\equiv& \!\! 
(
{2^{1/3}}/{\vartheta^{1/3}} \cdot \varepsilon
+
{\vartheta^{1/3}}/{2^{1/3}}
)/3, \quad
\theta \equiv -{2a^2}/{3}+m_0^2, \quad
\eta \equiv \sqrt{m_0^2-a^2}, \quad
\nonumber\\* 
\varepsilon \!\! &\equiv& \!\! 
a^4+24m_0 \, \ell_p^3, \quad
\vartheta \equiv 
2 a^6
+72 \,\delta \, \ell_p^3 
-2\sqrt{- \varepsilon^3 + (a^6+36 \delta \, \ell_p^3 )^2}, \quad 
\delta \equiv (3 m_0^2-2 a^2)m_0,
\nonumber
\end{eqnarray}
where 
we have obtained these 
by solving $\tD_4=0$ all at once 
using Mathematica, 
and then expanding the resulting solution.
However, it can be verified that
$r_{\pm}$ are the solutions 
when $r^2 - 2m_0 r + a^2=0$ is taken 
as the leading-order contribution, 
which agree with the $r_\pm$ in (\ref{btyn}).
On the other hand, $r_{\rm I \pm}$ are the solutions
when $a^2r^2$ and $2m_0 \ell_p^3$ are taken 
as the leading-order contribution.

The non-expanded $r_{\rm I\pm}$ and $r_\pm$ 
(the middle ones in (\ref{rtym})) exactly satisfy $\tD_4=0$, 
and the expanded $r_{\rm I\pm}$ and $r_\pm$ 
(the right ones in (\ref{rtym})) satisfy $\tD_4=0$ 
up to $\ell_p^3$ order.
Here, if the power expansion begins at $\ell_p^{3/2}$ order, 
$r_{\rm I\pm}$ must include terms up to $\ell_p^{9/2}$ 
in order for $\tD_4=0$ to be satisfied up to $\ell_p^3$ order.
In fact, the $\ell_p^{9/2}$-order term of the solution contributes 
to the $\ell_p^3$ order of the equation (\ref{ladv})\footnote{
$
\frac{2m_0 \ell_p^3}{r^2}
\big\vert_{
r = \gamma_1 \ell_p^{3/2} + \gamma_2 \ell_p^3 +\gamma_3 \ell_p^{9/2}
}
=
\frac{2m_0} {\gamma_1^2}
-\frac{4m_0 \gamma_2^3} {\gamma_1^3}\ell_p^{3/2}
+\frac{2m_0(3\gamma_2^2-2\gamma_1\gamma_3)}{\gamma_1^4}\ell_p^3+\cdots
$.
}.

$r_\pm$ in this Appendix also suffer from the same issue 
mentioned at the end of Sec.\,\ref{subChap:NHEK2}. 
Hence, we obtain $r$ in the extremal case 
from alternative conditions,
which are
$\tD_4 = {\cal O}(\ell_p^4)$ and
$(\tD_4)' = {\cal O}(\ell_p^4)$ 
under the assumption that
the forms of $r$ and $m_0$ are given by
$r=a+ \alpha \ell_p^3/a^3$ and 
$m=a+ \beta \ell_p^3/a^3$.
As a result, 
$r$ and $m_0$ in the extremal case are obtained as
$r=a(1+3\ell_p^3/a^3)$ and $m_0=a(1+\ell_p^3/a^3)$, respectively.
These agree with $r_{\rm ext}$ and $m_{\rm ext}$ in (\ref{rvsrwh}). 

\section{On the impossibility of determining the extremal values of $a$ and $m_0$ 
from the condition $r_+ = r_-$}
\label{App:kukll} 

We have mentioned the difficulty 
referred to in the title of this Appendix 
at the end of Sec.\,\ref{subChap:NHEK2} and Appendix\,\ref{App:Delta4}.
Actually, 
one might first consider using the condition $r_+=r_-$ 
to determine the extremal values of $a$ and $m_0$.
However, it turns out that 
the correct $a$ and $m_0$ cannot be obtained from this. 
In this Appendix, we discuss this issue.
\newline

We can obtain the following $a$ and $m_0$ 
from the condition $r_+=r_-$:
\begin{eqnarray}\label{rehj}
a = m_0 - \ell_p^3/{2m_0^2} \equiv a_\phi 
\quad \! \textrm{or} \quad \!
m_0 = a + \ell_p^3/{2a^2} \equiv m_\phi.
\end{eqnarray}
These $a_\phi $ and $m_\phi$ ultimately turn out to be incorrect.
With this in mind, we proceed with the discussion in this Appendix.
Using (\ref{rehj}), $r_\pm$ in the extremal limit are obtained as
\begin{subequations}\label{idsv}
\begin{align} 
\label{idsv1}
r_\pm \vert_{a=a_\phi+\alpha} 
=&\,\,
m_0 + {2\ell_p^3}/{m_0^2} +{\cal O}(\ell_p^{7/2}) \equiv r_\phi(m_0), \\ 
\label{idsv2}
r_\pm \vert_{m_0=m_\phi+\alpha}
=&\,\,
a + {5\ell_p^3}/{2a^2} +{\cal O}(\ell_p^{7/2}) \equiv r_\phi(a).
\end{align} 
\end{subequations}
However, these are not meaningful as locations of the horizon, 
because they do not satisfy $\tD = 0$ up to $\ell_p^3$ order. 
We show this in what follows.

We first assume the form of $r$ in $\tD$ in (\ref{RRBHTD})
as $\gamma_0+\gamma_1 \ell_p^3$.
As a result, $\tD$ is expressed as
\begin{eqnarray}\label{mvrle}
\tD\vert_{r=\gamma_0+\gamma_1 \ell_p^3} \!\! &=& \!\!\! 
f_0 + f_1 \ell_p^{3} + f_2 \ell_p^{6} + f_3 \ell_p^{9} + f_4 \ell_p^{12} + f_5 \ell_p^{15}, 
\end{eqnarray}
\vspace{-8mm}
\begin{eqnarray}
\hspace{-20.3mm}
{\rm where}\quad \! 
f_0 \!\! &=& \!\!\! 
\gamma_0^3 ( a^2-2 m_0 r_0+\gamma_0^2), \quad
f_1 = 
a^2 
+ (1+3 a^2 \gamma_1)\gamma_0^2 
+ \gamma_1 \gamma_0^3 
- 8 m_0 \gamma_1 \gamma_0^3 
+ 5 \gamma_1 \gamma_0^4, 
\nonumber\\
\hspace{-20.3mm}
f_2 \!\! &=& \!\!\! 
\gamma_1 \gamma_0 (2 +3 a^2 \gamma_1 -12 m_0 \gamma_1 \gamma_0+10 \gamma_1 \gamma_0^2), \quad
f_3 = 
\gamma_1^2 (
1
+a^2 \gamma_1
-8 m_0 \gamma_1 \gamma_0
+10 \gamma_1 \gamma_0^2 
), 
\nonumber\\
\hspace{-20.3mm}
f_4 \!\! &=& \!\!\! 
\gamma_1^4(5 \gamma_0 -2 m_0), \quad
f_5 = 
\gamma_1^5. 
\nonumber
\end{eqnarray}

Now, we take the extremal limit 
when $m_\phi$ and $a_\phi$ in (\ref{rehj}) are the extremal values of $m_0$ and $a$.
From (\ref{btyn}), 
we can know what we should substitute into $\gamma_0$ and $\gamma_1$, 
if we substitute the $r$ as the solution.
Therefore, we perform the following substitutions as the extremal limit:
\begin{subequations}\label{vebre}
\begin{align} 
\label{vebre1}
f_0 +f_1 \ell_p^3\,
\vert_{r_0 = m_0 \pm \eta,\, \gamma_1 
= \mp {m_0}/{\gamma_0^2 \eta}}\,
\vert_{m_0 = m_\phi,\,\eta = \eta_\phi(a_\phi)}
=& \,\,
{\cal O}(\ell_p^{7/2}), \\
\label{vebre2}
f_2 \ell_p^6\,
\vert_{r_0 = m_0 \pm \eta,\, \gamma_1 
= \mp {m_0}/{\gamma_0^2 \eta}}\,
\vert_{m_0 = m_\phi,\,\eta = \eta_\phi(a_\phi)}
=& \,\,
a_\phi^2 \ell_p^3+ {\cal O}(\ell_p^{7/2}), \\
\label{vebre3}
f_3 \ell_p^9\,
\vert_{r_0 = m_0 \pm \eta,\, \gamma_1 
= \mp {m_0}/{\gamma_0^2 \eta}}\,
\vert_{m_0 = m_\phi,\,\eta = \eta_\phi(a_\phi)}
=& \,\,
{\cal O}(\ell_p^{9/2}), \\
\label{vebre4}
f_4 \ell_p^{12}\,
\vert_{r_0 = m_0 \pm \eta,\, \gamma_1 
= \mp {m_0}/{\gamma_0^2 \eta}}\,
\vert_{m_0 = m_\phi,\,\eta = \eta_\phi(a_\phi)}
=& \,\,
{\cal O}(\ell_p^{6}), \\
\label{vebre5}
f_5 \ell_p^{15}\,
\vert_{r_0 = m_0 \pm \eta,\, \gamma_1 
= \mp {m_0}/{\gamma_0^2 \eta}}\,
\vert_{m_0 = m_\phi,\,\eta = \eta_\phi(a_\phi)}
=& \,\,
{\cal O}(\ell_p^{15/2}), 
\end{align} 
\end{subequations}
where
$\eta \vert_{m_0 = m_\phi} = a_\phi^{-1/2} \, \ell_p^{3/2} 
+{\cal O}(\ell_p^{7/2}) \equiv \eta_\phi(a_\phi)$
and 
$\eta \vert_{a = a_\phi} = m_o^{-1/2} \, \ell_p^{3/2} 
+{\cal O}(\ell_p^{7/2})\equiv \eta_\phi(m_\phi)$~($\eta$ is defined in (\ref{btyn})).
Accordingly, $\tD$ in the extremal limit is obtained as
\begin{subequations}\label{mjsb}
\begin{align} 
\label{mjsb1}
\tD \,
\vert_{r=\gamma_0+\gamma_1 l_p^3} \,
\vert_{r_0 = m_0 \pm \eta,\,\gamma_1 = \mp {m_0}/{\gamma_0^2 \eta}} \,
\vert_{m_0 = m_\phi,\,\eta = \eta_\phi(a_\phi)}
=& \,\, a_\phi^2 \ell_p^3+{\cal O}(\ell_p^{7/2}),\\
\label{mjsb2}
\tD \,
\vert_{r=r_0+r_1 \ell_p^3} \,
\vert_{r_0 = m_0 \pm \eta,\,r_1 = \mp {m_0}/{r_0^2 \eta}} \,
\vert_{a = a_\phi,\,\eta = \eta_\phi(a_\phi)}
=& \,\, m_\phi^2 \ell_p^3+{\cal O}(\ell_p^{7/2}).
\end{align} 
\end{subequations}
$r_\pm$ should satisfy $\tD=0$ up to $\ell_p^3$ order; 
however, they do not 
in the extremal limit by (\ref{rehj}),
as can be seen above. 

Its reason is that 
$\eta$ acquires $\ell_p^{3/2}$ dependence 
in the extremal limit 
derived from (\ref{rehj}),
as written after (\ref{vebre}),
Due to this, 
the term $f_2 \ell_p^{6}$ in (\ref{mvrle}), 
which is originally in the negligible order, 
shifts to $\ell_p^3$ order.
The origin of this fractional power $\ell_p^{3/2}$ is that 
we imposed the condition that
$r_+$ and $r_-$ coincide directly, 
as written above (\ref{rehj}).
In fact, one should analyze the tangent behavior at the extremal limit, 
as described in Sec.\,\ref{subChap:NHEK2} 
or at the end in Appendix\,\ref{App:Delta4}.
\newline

In the discussion above, 
we attempted to obtain the extremal case 
by directly substituting (\ref{rehj}), 
and this approach did not succeed.
Therefore, we consider softly taking the extremal limit as follows:
\begin{subequations}\label{hjyts}
\begin{align} 
r_\pm \vert_{m_0=m_\phi+\alpha}
=& \,\,
a \pm \sqrt{2 \alpha a_0}+{\cal O}(\alpha ^{3/2})
+\frac{1}{a_0^{3/2}}
\Big(
\mp \frac{1}{2 \sqrt{2\alpha}}
+\frac{5}{2a_0^{1/2}}
\mp\frac{35 \sqrt{\alpha}}{8 \sqrt{2}a_0}
+O(\alpha)
\Big) \ell_p^3+\cdots,
\\
r_\pm \vert_{a=a_\phi-\alpha}
=& \,\,
m_0 \pm \sqrt{2 \alpha m_0}+{\cal O}(\alpha^{3/2})
+\frac{1}{m_0^{3/2}}
\Big(
\mp\frac{1}{2 \sqrt{2\alpha}}
+\frac{2}{m_0^{1/2}}
\mp\frac{53 \sqrt{\alpha}}{8 \sqrt{2}m_0}
+O(\alpha)
\Big) \ell_p^3+\cdots, 
\end{align} 
\end{subequations}
where $\alpha$ is a parameter 
and $\alpha \to 0$ corresponds 
to softly taking the extremal limit.
Since these satisfy $\tD=0$ 
up to $\ell_p^3$ order 
even at $\alpha \to 0$, 
these are valid as the locations of the horizon.
However, they are not meaningful 
because their expressions break down as 
the terms in their expressions
diverge at $\alpha \to 0$. 

Therefore, 
we concluded that 
obtaining the extremal case 
from the condition $r_+=r_-$ does not go well, 
and we have finally obtained it 
by the way described in Sec.\,\ref{subChap:NHEK2} 
or at the end of Appendix\,\ref{App:Delta4}, 
which is consistent with the extremal case.
 
\section{Derivation of $t \tk \ll 1$ for $t \gg 1$}
\label{App:orhwl} 

In Sec.\,\ref{wryus}, 
the expression for $S_{\textrm{rad}}$ 
in the final stage of evaporation has been evaluated 
under the assumption that $t \tk \ll 1$ for $t \gg 1$. 
In fact, 
$t$ is large but finite, whereas $\tk$ approaches 0. 
Hence, we expect $t \tk \ll 1$ for $t \gg 1$.
However, this is not trivial.
Therefore, we examine this 
in this Appendix.
\newline

The energy emitted per unit time 
from a BH is generally given by
\begin{eqnarray}\label{ierf}
\frac{dE}{dt}
=\sum_{l,m}
\int^\infty_0 d\omega\,
\frac{\omega}{2\pi} 
\frac{\Gamma_{(\omega l m)}}{e^{\beta \omega}-1}, 
\end{eqnarray}
where $\Gamma_{(\omega l m)}$ are the greybody factors, 
and their expressions in the low-frequency limit, $m \omega \ll 1$, 
in the Schwarzschild BH are known to be
$\Gamma_{(\omega 00)}=\Gamma_{(\omega 0)}={{\cal A}\,\omega^2}/{\pi}=4 (\omega r_s)^2$ 
(${\cal A}$ and $r_s$ represent the horizon area and horizon radius, respectively. 
$\Gamma_{(\omega l)}$ for $l=1,2,3,\cdots$ are infinitesimal and negligible).

The expressions for the greybody factors 
in Kerr BH are also known in the low-frequency limit, 
and, strictly speaking, we should evaluate (\ref{ierf}) using these. 
However, this calculation is highly complicated.
Then, the purpose in considering (\ref{ierf}) in this Appendix is 
to examine whether $t\tk$ is damped or not in the extremal limit, 
for which exact calculation is not necessary. 
Therefore, we may perform the calculations in this Appendix
based on the result of (\ref{ierf}) evaluated in the Schwarzschild BH case.
In this case, since the calculations are an approximation, 
there is no meaning in performing the calculations 
up to $\ell^3$ order.
Therefore, we neglect the $\ell^3$ order corrections
in the calculations in this Appendix.

As such, 
(\ref{ierf}) is expressed in the Schwarzschild case as
\begin{eqnarray}\label{ewrt}
\frac{dE}{dt} \sim
\int^\infty_0 \frac{d\omega}{2\pi^2} 
\frac{{\cal A} \, \omega^3}{e^{\beta \omega}-1} 
= {\cal A}\,\beta^{-4}\, \Gamma(4) \,\zeta(4)
= \frac{\pi^2}{30}\,{\cal A} \,T_{\rm H}^4,
\end{eqnarray}
where $T_{\rm H} \equiv \beta^{-1}$.
$E$ and $\omega$ are quantities 
that do not include $G_{\rm N}$, 
defined in (\ref{ytvds})~(for this point, 
see Sec.\,\ref{subChap:perpe}). 
Hence, their dimension is energy, which are -1 in units of length, 
and the dimensions of both sides agree with each other.
Looking at (\ref{ewrt}), 
integrating $\omega$ up to $\infty$ may appear 
to contradict the assumption of the low-frequency limit. 
However, 
the high-frequency contributions are exponentially suppressed, 
and it causes no problem. 
Actually, the result above is consistent with the Stefan-Boltzmann law, 
${dE}/{dt} \propto T_{\rm H}^4$.

In the Schwarzschild case, 
${\cal A}=4\pi (2m_0)^2$ and 
$T_{\rm H}=1/8 \pi m_0$~(these $m_0$ include $G_{\rm N}$).
Applying these to (\ref{ewrt}), 
we obtain the following result:
\begin{eqnarray}\label{ypee}
\frac{dm_0}{dt}=-\frac{1}{7680\pi}\frac{G_{\rm N}}{m_0^2} 
\,\,\Rightarrow\,\,
m_0(t)=m_0(1-\frac{G_{\rm N} t}{2560 m_0^3\pi})^{1/3}, 
\end{eqnarray}
where we have identified $E$ with $m_0$.
This $m_0$ includes $G_{\rm N}$, 
while $E$ does not 
as mentioned after (\ref{ewrt});~consequently, 
$G_{\rm N}$ has appeared in the r.h.s.. 
The dimensions of both sides agree with each other.

Using (\ref{ypee}), 
$\tk^{\rm (non-ext)}$ in (\ref{thyrthrt}) is expressed as
\begin{eqnarray} \label{prvr}
\tk^{\rm (non-ext)}=
\frac{1}{2}(
\frac{1}{m_0 \, x}
+\frac{-m_0x+\sqrt{m_0^2 \, x^2-a^2}}{a^2}
), \quad
x \equiv (1-\frac{G_{\rm N} t}{2560 m_0^3\pi})^{1/3},
\end{eqnarray} 
where, before we substitute (\ref{ypee}), 
we expressed $\tk^{\rm (non-ext)}$ 
in terms of $a$ and $m_0$ 
by substituting $r_\pm$ in (\ref{btyn}).
As shown in Sec.\,\ref{Chap:rgtr},
$\tk^{\rm (non-ext)}$ vanishes at the extremal limit.
Therefore, from $(\ref{prvr})=0$, 
the upper bound of $t$ is determined, 
and its range is obtained as 
\begin{eqnarray}\label{tory}
0 \le t \le t^*, \quad 
t^* \equiv 2560\pi m_0^3\,(1-(a/m_0)^3)/G_{\rm N}.
\end{eqnarray} 

\begin{figure}[H] 
\vspace{-11.0mm} 
\begin{center}
\includegraphics[clip,width=8.7cm,angle=-90]{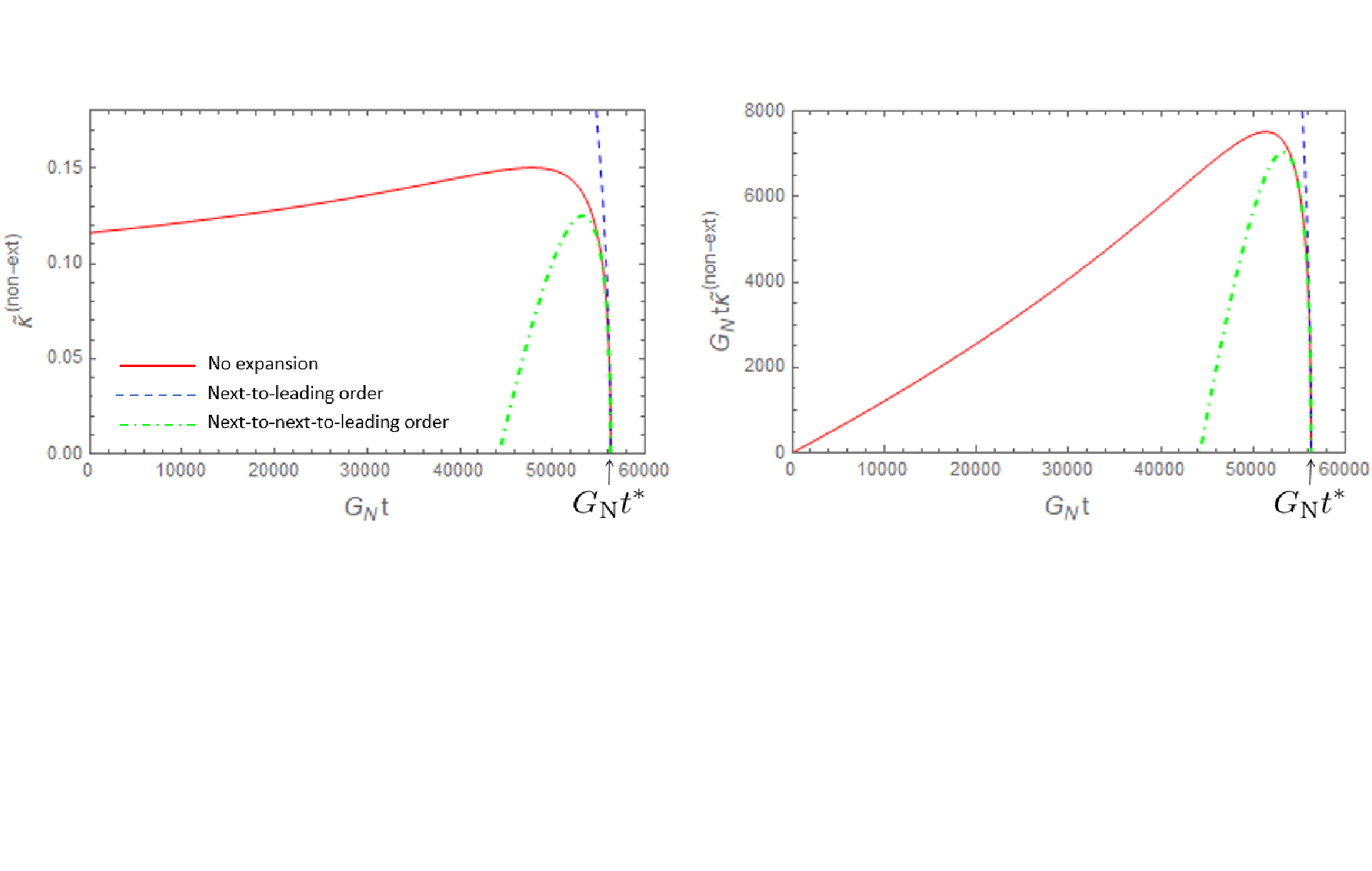} 
\end{center}
\vspace{-40.0mm}
\caption{
``No expansion''  is obtained from (\ref{prvr}).
On the other hand, ``Next-to-leading order'' and ``Next-to-next-to-leading order'' 
are obtained 
by retaining only the first term and the first two terms in (\ref{trjtyk1}), respectively.
$t^*$ is defined in (\ref{tory}), and $G_{\rm N}t^* = {\cal}O(G_{\rm N}^0)$.
The same applies to the right figure for (\ref{trjtyk2}).
$a=1$ and $m_0=2$ in these plots.
Note that the horizontal axis represents $G_{\rm N}t$ 
and the vertical axis in the right figure represents $G_{\rm N} t\tk^{\rm (non-ext)}$.
} 
\label{rjrnd}
\end{figure}

We examine the behaviors of $\tk^{\rm (non-ext)}$ and $t \tk^{\rm (non-ext)}$ 
near the extremal point.
For this purpose, we express $x$ in the expansion form around $t=t^*$ as 
\begin{eqnarray}
x =
{a}/{m_0}
+{G_{\rm N}\delta t}/{7680a^2m_0\pi}
-{G_{\rm N}^2\delta t^2}/{58982400a^5m_0 \pi^2}
+{\cal O}(\delta t^3), \quad \delta t \equiv t^*-t.
\end{eqnarray}
Using this, we can obtain the following results:
\begin{subequations}\label{trjtyk}
\begin{align} 
\label{trjtyk1}
\cdot \quad \hspace{1.4mm}
\tk^{\rm (non-ext)}
\vert_{t\sim t^*}&=
\frac{1}{32}\sqrt{\frac{G_{\rm N}\delta t}{15a^5\pi}}
+\frac{G_{\rm N}\delta t}{7680 a^4\pi} 
+{\cal O}(\delta t^{3/2}),
\\ 
\label{trjtyk2}
\cdot \quad
t\tk^{\rm (non-ext)}
\vert_{t\sim t^*}&=
\frac{16(m_0^3-a^3)}{a^{5/2}}
\sqrt{\frac{5 \pi }{3}\,\frac{\delta t}{G_{\rm N}}} 
+\frac{a^3-m_0^3}{3 a^4}\,\delta t
+{\cal O}(\delta t^{3/2}),
\end{align} 
\end{subequations}
where 
$
G_{\rm N}\delta t
=G_{\rm N}t^*-G_{\rm N}t
=2560\pi m_0^3\,(1-(a/m_0)^3)-G_{\rm N}t
$ and
$
\delta t
=(G_{\rm N}t^*-G_{\rm N}t)/G_{\rm N}
=(2560\pi m_0^3\,(1-(a/m_0)^3)-G_{\rm N}t)/G_{\rm N}
$.
We plot $\tk^{\rm (non-ext)}$ and $G_{\rm N} t\tk^{\rm (non-ext)}$ for $G_{\rm N} t$
in Fig.~\ref{rjrnd},
from which we observe that
$t \tk$ reaches $0$ 
as $t \to t^*$.

\section{$r_b$ determined from the extremal condition and $\tk r_b$}
\label{ryuyj}

In this appendix, 
we obtain $r_b$ in $S_{\textrm{rad}}^{( \textrm{late})}$ in (\ref{yethte}) 
from 
$\partial S_{\textrm{rad}}^{( \textrm{late})}/ \partial r_b =0$, 
and show the value of $\tk r_b$ by that $r_b$ near $\alpha=0$.

\subsection{The extremum of $r_b$}
\label{ryuyj1}

\begin{figure}[H] 
\vspace{-16.0mm} 
\begin{center}
\hspace{-0.0mm} 
\includegraphics[clip,width=11.0cm,angle=-90]{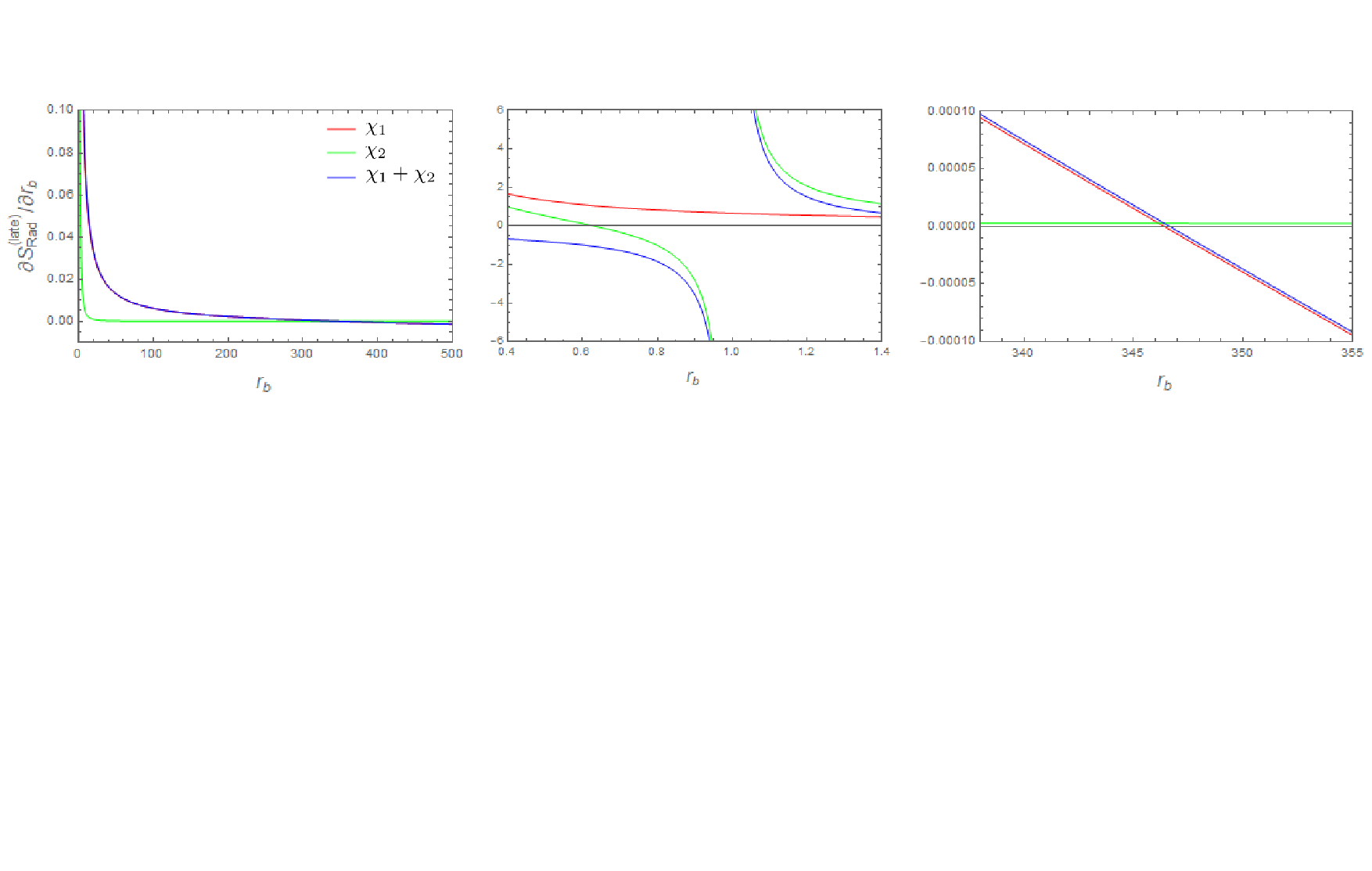}  
\end{center}
\vspace{-67.0mm}
\caption{These are plots of $\partial S_{\textrm{rad}}^{( \textrm{late})}/ \partial r_b$ 
in (\ref{gtrhs}) for $\alpha$.
The left figure shows the overall behavior, 
while the middle and right figures show 
its small and large $r_b$ regions. 
It can be seen in the middle and right figures that 
there is a stationary point of $S_{\textrm{rad}}^{( \textrm{late})}$ 
in each of the small $r_b$ and large $r_b$ regions.}
\label{ygiul}
\end{figure}

First, 
$
\partial S_{\textrm{rad}}^{(\textrm{late})} / \partial r_b$ 
is calculated as follows:
\begin{align}\label{gtrhs}
\frac{\partial S_{\textrm{rad}}^{(\textrm{late})}}{\partial r_b}
=& \,\,\,
\frac{1}{3}
\frac{\partial}{\partial r_b}
\ln 
\Big\vert
\frac{1-\cosh(\tk r_b)}{1+\cosh (\tk r_b)}
\Big\vert
+\frac{1}{6}\frac{\partial}{\partial r_b}
\ln \big[\tf(r_b) \big]
\nonumber\\*
=& \,\,
\underbrace{
\frac{2}{3r_b}
-\frac{r_b \alpha}{18a^3}}_{\equiv \,\chi_1}
\underbrace{
-\frac{a(a+r_b)}{3(a-r_b)(a^2+r_b^2)}
-\frac{2a\alpha}{3(a-r_b)^3}}_{\equiv \,\chi_2},
\end{align}
where $\chi_1$ and $\chi_2$ originate 
from the second and third terms 
in the first line in (\ref{yethte}), respectively.
We show the behavior of (\ref{gtrhs}) for $r_b$ in Fig.\,\ref{ygiul}.
From this, 
we find that 
there are two stationary points
in the small $r_b$ and the large $r_b$ regions, respectively\footnote{
The origin of these two stationary points is unclear. 
The stationary point in the small $r_b$ region may be 
an artifact arising from the present analysis.
The same can also be true for the result in (\ref{uygu})}.

Since the cutoff surface is generally assumed to be located far away, 
we focus on the solution 
in the large $r_b$ region.
Then, we can see from the right figure in Fig.\,\ref{ygiul} that
the behavior of (\ref{gtrhs}) in the large $r_b$ region
is mostly determined 
by $\chi_1$.
Therefore, the extremum $r_b$ is determined as
\begin{align}\label{rrrst}
\chi_1=0 
\quad \!\! \Rightarrow \quad \!\!
r_b=2\sqrt{3}\,a^{3/2}/\sqrt{\alpha}.
\end{align}

\subsection{The behavior of $\tk r_b$ and the expression of $S_{\textrm{rad}}^{(\textrm{late})}$}
\label{wryus3}

With $r_b$ in (\ref{rrrst}), 
$\tk r_b$ behaves near $\alpha = 0$ as follows:
\begin{align}\label{fyur}
\tk r_b= 
\sqrt{6}
-2\sqrt{3\alpha/a}
+(
\sqrt{6}/a^3
-{14\sqrt{3\alpha}}/{a^{7/2}}
)\ell_p^3
+{\cal O}(\ell_p^4)
+\cdots,
\end{align}
where ``$\cdots$'' represnets the higher order terms with regard to $\alpha$.
Substituting (\ref{fyur}) into $S_{\textrm{rad}}\big\vert_{t_a,t_b \gg 1}$ in (\ref{yethte}), 
we obtain the following result: 
\begin{eqnarray}\label{ywere}
&& 
S_{\textrm{rad}}\big\vert_{t_a,t_b \gg 1}\big\vert_{\rm (\ref{fyur})}
=
\frac{\cal A}{2G_{\rm N}} 
+
\frac{1}{6} \ln 
\big\vert
(\frac{2}{\epsilon\tk})^4 \,
\tf( r_a ) 
\big\vert
+\frac{1}{3}
\ln 
\Big\vert
\frac{1-\cosh (\tk \, r_b)}{1+\cosh (\tk r_b)} 
\Big\vert
\bigg\vert_{\rm (\ref{fyur})},
\\*
&& 
\frac{1}{3}
\ln 
\Big\vert
\frac{1-\cosh (\tk r_b)}{1+\cosh (\tk r_b)} 
\Big\vert
\bigg\vert_{\rm (\ref{fyur})}
\! =
-4 \sqrt{\frac{3\alpha}{a}} \frac{1}{\sinh(\sqrt{6})}
+2 \ln[\tanh [\sqrt{\frac{3}{2}}]]
+\frac{2 \sqrt{6} }{a^3\sinh(\sqrt{6})}\ell_p^3
+\cdots, 
\nonumber
\end{eqnarray}
where $\tf( r_b ) \approx 1$ in  $\alpha \to 0$~($\tf(r)$ is defined in (\ref{MetricsType})). 


\end{document}